\documentclass[reprint,showpacs,
aps]{revtex4-1}

\usepackage{amsmath, amssymb}
\usepackage{type1cm}
\usepackage[dvipdfmx]{color}
\usepackage[dvipdfmx]{graphicx}
\usepackage{dcolumn}
\usepackage{bm}

\newcommand{\ii}{\text{i}}

\begin{document}

\preprint{APS/123-QED}

\title{Supersymmetry breaking and Nambu-Goldstone fermions with cubic dispersion} 

\author{Noriaki Sannomiya$^1$\thanks{E-mail address: sannomiya@cams.phys.s.u-tokyo.ac.jp}, Hosho Katsura$^1$, and Yu Nakayama$^{2,3}$}

\affiliation{
$^1$Department of Physics, Graduate School of Science, The University of Tokyo, Hongo, Tokyo 113-0033, Japan\\
$^2$Department of Physics, Rikkyo University, Toshima, Tokyo 177-8501, Japan\\
$^3$Kavli Institute for the Physics and Mathematics of the Universe (WPI),
University of Tokyo, 5-1-5 Kashiwanoha, Kashiwa, Chiba 277-8583, Japan}

\date{\today}
\begin{abstract}
We introduce a lattice fermion model in one spatial dimension with supersymmetry (SUSY) but without particle number conservation. 
The Hamiltonian is defined as the anticommutator of two nilpotent supercharges $Q$ and $Q^\dagger$. Each supercharge is built solely from spinless fermion operators and depends on a parameter $g$. The system is strongly interacting for small $g$, and in the extreme limit $g=0$, the number of zero-energy ground states grows exponentially with the system size. By contrast, in the large-$g$ limit, the system is non-interacting and SUSY is broken spontaneously. We study the model for modest values of $g$ and show that under certain conditions spontaneous SUSY breaking occurs in both finite and infinite chains. We analyze the low-energy excitations both analytically and numerically. Our analysis suggests that the Nambu-Goldstone fermions accompanying the spontaneous SUSY breaking have cubic dispersion at low energies.
\end{abstract}

\maketitle


\section{\label{sec:level1}Introduction}
Besides an artificial fine-tuning, one non-trivial way to guarantee the existence of gapless excitations is to use spontaneous symmetry breaking. If ordinary bosonic symmetry is spontaneously broken, we expect the emergence of gapless bosonic degrees of freedom known as Nambu-Goldstone (NG) bosons. If, on the other hand, fermionic symmetry is spontaneously broken, we expect the emergence of gapless fermionic degrees of freedom known as NG fermions.

One important example of fermionic symmetries is what is called supersymmetry  (SUSY)~\cite{Wess_NPB74,Witten_NPB82}, where the anticommutator of its generators gives the Hamiltonian. A motivation of SUSY in particle physics is to render the hierarchy problem less severe, but SUSY itself has yet to be observed experimentally. Therefore, SUSY, if any, must be spontaneously broken in our world. In relativistic systems, it is known that spontaneous SUSY breaking leads to massless fermions called Goldstinos as a NG fermion~\cite{Witten_NPB82, Salam:1974zb}. 
An effective description of the Goldstinos is well-understood by using the non-linear realization of the SUSY~\cite{PLB_Volkov}. 

On the other hand, explicit SUSY in lattice models~\cite{PRL_Fendley_2003, PRL_Fendley_2005, PRL_Yu, PRL_Huijse, EPJB_huijse, PRL_Yu_2010, NJP_Huijse} and emergent SUSY at quantum critical points~\cite{PRB_Lee_2007, Science_Grover, PRL_Jian,PRL_Rahmani,arXiv_Li,arXiv_Jian} have recently been discussed  in the condensed matter literature. 
In some systems, spontaneous SUSY breaking occurs and gives rise to NG fermions. For example, Yu and Yang  introduced a model with SUSY in the context of cold atoms. In the model, it was shown that SUSY is spontaneously broken and there exist gapless excitations with quadratic dispersion, which implies the existence of NG fermions \cite{PRL_Yu}. In $2+1$ and $3+1$ dimensional topological superconductors, the topologically protected edge/surface Majorana fermion is identified with a NG fermion arising from spontaneous SUSY breaking \cite{Science_Grover}.

In contrast to relativistic systems, the nature of NG fermions in non-relativistic systems is less well-understood. While the classification theory of non-relativistic NG bosons~\cite{PR_Nambu,NC_Goldstone,PR_Goldstone} has received renewed attention \cite{PRL_Watanabe,PRL_Hidaka}, its naive application to SUSY may lead to a wrong conclusion. In our previous work, we developed a theory of NG fermions in lattice systems and studied the properties of NG fermions in the extended Nicolai model~\cite{U1Nicolai}, which is a generalization of the model studied by Nicolai in 1970s~\cite{JPA_Nicolai_76, JPA_Nicolai_77}. The model exhibits spontaneous SUSY breaking accompanied by NG fermions with a linear dispersion. We have clarified at which point, the hidden assumption in the argument of the NG bosons such as decoupling from the other gapless excitations, is violated.

In this paper, we introduce another curious example of spontaneous SUSY breaking. The model is constructed solely out of fermions on the lattice, which is analogous to the Nicolai model, but it has only $\mathbb{Z}_2$ symmetry rather than the ${\rm U}(1)$ symmetry. It turns out that the NG fermions have a cubic dispersion relation in the wave number $p$ without fine-tuning. This is again unexpected from the general theories of non-relativistic NG bosons. We note that the model with cubic dispersion is discussed in the study of topological phases of matter \cite{JETP_Volovik,PRB_Guinea,Zunger_arxiv} and quantum spin liquids \cite{PRB_Biswas, PRB_Wang}, and our model may be used to naturally realize such a cubic dispersion relation. 

The organization of this paper is as follows. In Sec. \ref{sec:model}, we introduce the system we study and describe the symmetries of the model. We then introduce the Hamiltonian as the anticommutator of the two supercharges. 
In Sec. \ref{sec:singlet}, we focus on the case where SUSY is unbroken and show that the number of SUSY singlets, the zero-energy ground states, grows exponentially with the system size. 
In Sec. \ref{sec:break}, we first provide a precise definition of spontaneous SUSY breaking. We then prove that  SUSY is spontaneously broken in finite and infinite chains when $g>0$ and $g>4/\pi$, respectively. 
In Sec. \ref{sec:NGf}, we study the low-energy properties of the model using rigorous inequality and numerical diagonalization. We provide strong evidence for the existence of a massless excitation and show that its dispersion is cubic. The conclusion is given in Sec. \ref{sec:conclusion}. 
In Appendix \ref{sec:openZES}, we present the results for the number of the ground states in the $g=0$ model with open boundary conditions. In Appendix \ref{sec:SYK}, we present a random generalization of the $\mathbb{Z}_2$ Nicolai model i.e. the SUSY Sachdev-Ye-Kitaev (SYK) model \cite{Gross_arXiv, Maldacena_arXiv}. The derivation of some of the formulas used in the main text is presented in Appendices \ref{sec:free} and \ref{sec:fdp}. In Appendix \ref{sec:ModifyQ}, we discuss the stability of the cubic dispersion against SUSY-preserving perturbations. In Appendix \ref{sec:2dSUSY}, we present a generalization of the model on a two-dimensional triangular lattice.

\section{Model}
\label{sec:model}
We consider a system of spinless fermions on a chain of length $N$. For each site $j$, we define a creation (annihilation) operator as $c_j^\dagger$ ($c_j$). These operators obey the canonical anticommutation relations
\begin{align}
\{c_j,c_i^\dagger\}=\delta_{j,i} \ , \ \{c_i,c_j\}=\{c_i^\dagger,c_j^\dagger\}=0,
\end{align}
for all $i,j=1,\dots,N$. We denote the number operators by $n_j:=c_j^\dagger c_j$ ($j=1,...,N$) and write the total fermion number as $F:=\sum_{j=1}^N n_j$.

\bigskip

\subsection{Supercharges and Hamiltonian}
\label{sec:Ham}
Let us define the supercharge as
\begin{align}
Q:=\sum_{j=1}^N(gc_j+c_{j-1}c_jc_{j+1}),
\label{eq:SuQ}
\end{align}
where periodic boundary conditions (PBC) are assumed. The other supercharge $Q^\dagger$ is defined as the Hermitian conjugate of $Q$. Both $Q$ and $Q^\dagger$ are nilpotent and made up solely of fermion operators. 
In terms of these supercharges, the Hamiltonian of our model is defined as 
\begin{align}
H=\{Q,Q^\dagger\}.
\label{eq:Ham}
\end{align}
We refer to this model as the $\mathbb{Z}_2$ Nicolai model. In the following, we consider the case $g\ge0$ since $H$ with $g\le0$ can be achieved by a local unitary transformation $\mathcal{C}$: $c_j\to \mathrm{i}c_j$, $c_j^\dagger \to -\mathrm{i}c_j^\dagger$. 
We frequently use the following properties that follow directly from the definition Eq. (\ref{eq:Ham}): (i) all energy eigenvalues of $H$ are non-negative, (ii) states with a positive energy come in pairs, and (iii) any state with zero energy is a ground state and is annihilated by both $Q$ and $Q^\dagger$~\cite{Witten_NPB82, PRL_Fendley_2003, PRL_Fendley_2005}.

\begin{figure}[htb]
\includegraphics[width=0.95\columnwidth]{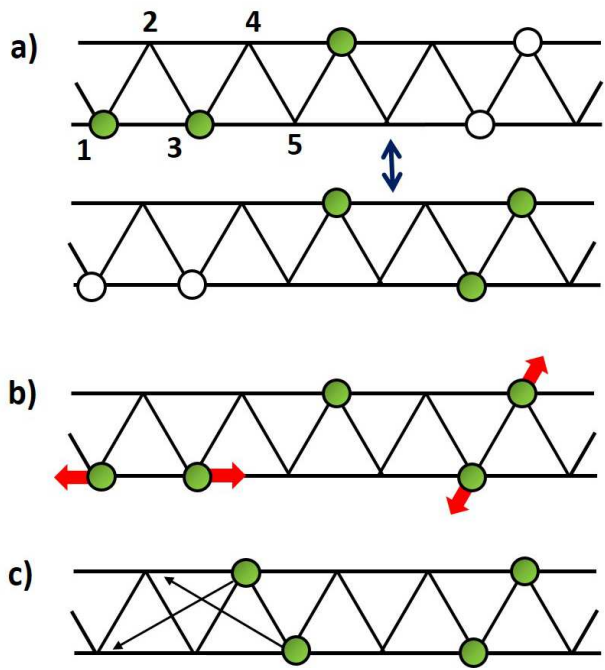}
\caption{Schematics of individual terms in the Hamiltonian $H$. (a) the paring term ($H_{\rm free}$), (b) the nearest-neighbor and the next-nearest-neighbor repulsive interactions ($H_1$), and (c) the first line of the third term ($H_2$). Green (gray) balls represent spinless fermions.}
\label{fig:schematic}
\end{figure}

Since each summand in $Q$ is local and fermionic, the Hamiltonian $H$ is local as well. To see this, let us derive the explicit expression for $H$. After some algebra, we have
\begin{align}
H=H_{\rm free}+H_{1}+H_{2}+g^2N,
\label{eq:hamz2}
\end{align}
where
\begin{align}
H_{\rm free} & =g\sum_{j=1}^N (2c_jc_{j+1} -c_{j-1}c_{j+1} +{\rm H.c.}), \label{eq:freeham} \\
H_{1} & =\sum_{j=1}^N(1-3n_j+2n_{j}n_{j+1}+n_{j}n_{j+2}), \label{eq:chargeham} \\
H_{2} & =\sum_{j=1}^N\left(c_j^\dagger c_{j-1}^\dagger c_{j+2}c_{j+3}+{\rm H.c.}\right)
\nonumber \\ 
& +\sum_{j=1}^N\left[(n_{j-1}+n_j-1)c_{j+1}^\dagger c_{j-2}+{\rm H.c.}\right].
\label{eq:pairham}
\end{align}
A schematic of each term in the Hamiltonian is shown in Fig. \ref{fig:schematic}. The first term $H_{\rm free}$ describes the pairing terms of nearest and next-nearest neighbor particles. 
Since it is quadratic, one can easily solve it (see Appendix \ref{sec:free} for details). 
The second term $H_{1}$ consists of the on-site potential and the repulsive interaction between two particles on nearest-neighbor or next-nearest-neighbor sites. 
The third term $H_{2}$ is rather complicated, but the first line represents a pair hopping term. The second line of Eq. (\ref{eq:pairham}) can be thought of as the third-neighbor hopping term, the amplitude of which is influenced by the presence or absence of fermions between the sites. 
 
\subsection{Symmetries}
\label{sec:sym}
The $\mathbb{Z}_2$ Nicolai model has various symmetries, including both fermionic and ordinary symmetries. 
The supercharges are conserved charges that commute with the Hamiltonian
\begin{align}
[H,Q]=[H,Q^\dagger]=0, 
\end{align}
which follows from the nilpotency of $Q$ and $Q^\dagger$. 
Because of the pairing terms, the model does not have U(1) symmetry. Instead, it has $\mathbb{Z}_2$ symmetry, i.e., $H$ commutes with the fermionic parity
\begin{align}
[H,(-1)^F]=0.
\end{align}
We note in passing that a supersymmetric lattice model with only ${\mathbb Z}_2$ symmetry but different from our $\mathbb{Z}_2$ Nicolai model has been studied in the context of integrable models \cite{JSM_Gier}.

We remark that the symmetry is enhanced to U($1$) when $g=0$, which follows from the fact that $Q$/$Q^\dagger$ at the point decreases/increases the fermion number $F$ by exactly three. As a result, the terms $H_1$ and $H_2$ in Eqs. (\ref{eq:chargeham}) and (\ref{eq:pairham}) commute with $F$.

Let us discuss the other symmetries. 
The supercharge $Q$ ($Q^\dagger$) is invariant under sending $c_j\to c_{j+1}$ ($c^\dagger_j \to c^\dagger_{j+1}$). 
As a consequence, the Hamiltonian is invariant under translation by one site. Next, we introduce the inversion-like unitary operator $U$ that acts as
\begin{align}
U^{-1}c_jU=\left\{ \begin{array}{ll}
\mathrm{i}c_{N-j} & j=1,\dots,N-1 \\
\mathrm{i}c_N & j=N.
\end{array} \right.
\label{eq:inversion_like}
\end{align} 
The explicit expression for $U$ in terms of fermion operators can be derived by noting that the operator $P_{i,j} = 1- (c^\dagger_i - c^\dagger_j) (c_i - c_j)$ permutes $c_i$ and $c_j$~\cite{Essler_Hubbard}.
Under this symmetry operation, the supercharges $Q$ and $Q^\dagger$ are invariant up to phase factors:
\begin{align}
U^{-1}QU=\mathrm{i}Q, \quad \ U^{-1}Q^\dagger U=-\mathrm{i}Q^\dagger.
\end{align}
Therefore, we have 
\begin{align}
U^{-1}HU= \{ U^{-1} Q U, U^{-1}Q^\dagger U \}
=H,
\end{align}
which implies that $U$ commutes with $H$. 
The operator $U$ plays a crucial role in our analysis below.

\bigskip

\section{SUSY singlets}
\label{sec:singlet}

In this section, we restrict ourselves to the case $g=0$ where the supercharge in Eq. (\ref{eq:SuQ}) becomes
\begin{align}
Q=\sum_{j=1}^Nc_jc_{j+1}c_{j+2},
\end{align}
and the Hamiltonian reduces to $H=H_1+H_2$. Similarly to the original Nicolai model~\cite{JPA_Nicolai_76, Moriya_arXiv}, SUSY is unbroken in this case, i.e., the ground-state energy is exactly zero. This is easily verified by noting that the state with a fermion on every other site,
\begin{equation}
\cdots \bullet \circ \bullet \circ \bullet \circ \bullet \circ \bullet \circ \bullet \circ \bullet \circ \bullet \circ \bullet \circ \cdots 
\nonumber
\end{equation}
is annihilated by both $Q$ and $Q^\dagger$. Here, $\circ$ and $\bullet$ denote empty and occupied sites, respectively, and the total number of sites $N$ is assumed to be even for simplicity. There are many other states annihilated by both $Q$ and $Q^\dagger$. For example, the state with a fermion on every third site, 
\begin{equation}
\cdots \bullet \circ \circ \bullet \circ \circ \bullet \circ \circ \bullet \circ \circ \bullet \circ \circ \bullet \circ \circ \cdots 
\nonumber
\end{equation}
is another ground state. In fact, the number of the zero-energy ground states grows exponentially with the system
 size. In Sec. \ref{sec:num_counting}, we present numerical results and discuss how fast the ground-state degeneracy increases with $N$. In Sec. \ref{sec:product_state}, using a transfer matrix method, we count the number of ground states which can be written as product states. In Sec. \ref{sec:Witten_ind}, we obtain a lower bound on the number of the ground states by computing the Witten index. 

\begin{widetext}
\begin{center}
\begin{table}[htb]
\begingroup
\renewcommand{\arraystretch}{1.3}
\begin{tabular}{ccccccccccccccc}
\hline \hline
~~~~~$N$~~~~~ & \,~3~\, & 4 & 5 & 6 & 7 & 8 & 9 & 10 & 11 & 12 & 13 & 14 & 15 & 16  \\ \hline
~~~~~$Z$~~~~~ & \,~6~\,  & ~12~ & ~20~ & ~36~ & ~54~ & ~108~ & ~172~ & ~324~ & ~530~ & ~984~ & ~1672~ & ~3028~ & ~5232~ & ~9388~
\\
~~~~$Z_{\rm cl}$~~~~ & ~6~ & ~6~ & ~10~ & ~20~ & ~28~ & ~46~ & ~78~ & ~122~ & ~198~ & ~324~ & ~520~ & ~842~ & ~1366~ & ~2206~
\\
~~~~~$W$~~~~~ & \,~6~\, & ~12~ & ~18~ & ~36~ & ~54~ & ~108~ & ~162~ & ~324~ & ~486~ & ~972~ & ~1458~ & ~2916~ & ~4374~ & ~8748~
\\
\hline \hline 
\end{tabular}
\endgroup
\caption{Ground-state degeneracy of the periodic chain with $g=0$ up to $N=16$ sites. $Z$, $Z_{\rm cl}$, and $W$ refer to the number of zero-energy states, the number of classical ground states, and the Witten index, respectively. }
\label{tab:ZES}
\end{table} 
\end{center}
\end{widetext}

\subsection{Numerical results}
\label{sec:num_counting}

The second row of Table \ref{tab:ZES} shows the number of the zero-energy ground states for periodic chains up to $N=16$ sites. The results are obtained by numerical diagonalization. The data obtained suggest that the number of the ground states ($Z$) grows exponentially with $N$. From the fit to the data, we find 
\begin{align}
Z \sim1.761^N,
\end{align}
where the data for $N=14$, $15$, $16$ are used for the fit. 
Thus we have an extensive ground-state entropy when $g=0$. The ground-state entropy per site reads $S_{\rm GS}/N=\ln Z/N \sim 0.566$. 

The model with open boundary conditions can be studied in the same way. From the results obtained, we conjecture that the ground-state entropy per site is exactly given by $S_{\rm GS}/N = \ln x^* \sim 0.571$, where $x^*$ is the real root of the cubic equation $x^3-2x-2=0$. The result obtained is remarkably close to the value in the periodic case. We provide evidence for this conjecture in Appendix \ref{sec:openZES}. 

\subsection{Counting classical ground states}
\label{sec:product_state}

An extensive ground-state entropy is a common feature in many supersymmetric lattice models and is called {\it superfrustration} \cite{PRL_Fendley_2005, PRL_Huijse, EPJB_huijse, NJP_Huijse}. 
To get a better understanding of superfrustration in our model, it is instructive to consider the product states annihilated by both $Q$ and $Q^\dagger$. In the following, they are referred to as {\it classical} ground states. In fact, they are the ground states of $H_1$, the classical part of the Hamiltonian where frustration in the classical sense exists because the nearest and next-nearest neighbor interactions cannot be minimized simultaneously. 

It is easy to see that a product state which does not contain the configuration $~\circ \circ \circ~$ or $~\bullet \bullet \bullet~$ in any three consecutive sites is a classical ground state. This can be rephrased as follows: a classical ground state is a state in which the configuration of any three consecutive sites is one of the following: $\{~\circ \circ \bullet, ~\circ \bullet \circ, ~\circ \bullet \bullet, ~\bullet \circ \circ, ~\bullet \circ \bullet, ~\bullet \bullet \circ~ \}$.  From this $3$-site rule, one can construct a transfer matrix and count the number of classical ground states exactly. Following the standard procedure, one can express $Z_{\rm cl}$, the number of classical ground states, in terms of the transfer matrix as $Z_{\rm cl} = {\rm Tr}\, T^N$, where
\begin{equation}
T = \left(
\begin{array}{cccc}
0 & 0 & 1 & 0 \\
1 & 0 & 1 & 0 \\
0 & 1 & 0 & 1 \\
0 & 1 & 0 & 0
\end{array}
\right).
\end{equation}
Here, the order of the basis states is $\{ ~\circ \circ, ~\circ \bullet, ~\bullet \circ, ~\bullet\bullet ~ \}$. The eigenvalues of $T$ can be computed analytically, and are given by
\begin{equation}
\lambda=\frac{1\pm \sqrt{5}}{2}, \quad \exp \left( \pm \frac{2\pi \bf i}{3} \right).
\end{equation}
Thus, we have
\begin{equation}
Z_{\rm cl} = \left( \frac{1+ \sqrt{5}}{2} \right)^N + \left( \frac{1- \sqrt{5}}{2} \right)^N +2 \cos \left( \frac{2\pi N}{3} \right).
\end{equation}
When $N$ is large, $Z_{\rm cl}$ is dominated by the contribution from the largest eigenvalue $\lambda_{\rm max}$ of $T$. Therefore, we have $Z_{\rm cl} = (\lambda_{\rm max})^N \sim (1.618)^N$ for large $N$. The fact that $Z_{\rm cl} < Z$ clearly shows the existence of entangled ground states that cannot be simply expressed as product states.

\subsection{Witten index}
\label{sec:Witten_ind}
We now derive a better lower bound for $Z$ by computing the Witten index. To this end, let us first consider the structure of the space of states. Because of the U($1$) symmetry at $g=0$, the fermion number $F$ is conserved and the Hamiltonian is block-diagonal with respect to $F$. Since  $Q$/$Q^\dagger$ decreases/increases $F$ by exactly three, the total Hilbert space can be divided into three sectors ${\cal H}_f$ ($f=0,1,2$), where ${\cal H}_f$ denotes the sector of all states with $F=f$ mod $3$. In each sector, the Witten index is defined as 
\begin{align}
W_f={\rm Tr}_{{\cal H}_f} [(-1)^F e^{-\beta H}],
\end{align}
where $\beta \ge 0$. 

Since all positive-energy states come in pairs with the same energy but the opposite $(-1)^F$, only the zero-energy states contribute to $W_f$. Therefore, $W=\sum_f |W_f|$ gives a lower bound for $Z$. The explicit value of $W_f$ can be computed by noting that each $W_f$ is independent of $\beta$ and can be evaluated in the limit $\beta \to 0$. After some manipulation of binomial coefficients, we have
\begin{equation}
W = \sum^2_{f=0} |W_f| = \left\{
\begin{array}{cc}
2 \times 3^{\frac{N-1}{2}}~~ & ~~N:\, {\rm odd}~~ \\
4 \times 3^{\frac{N}{2}-1}~ & ~~N:\, {\rm even}~
\end{array}
\right. .
\end{equation}
For both even and odd $N$, the Witten index $W$ grows exponentially with $N$ and a lower bound for the ground-state entropy per site is obtained as $S_{\rm GS}/N \ge \ln 3/2 = 0.549...$, which is slightly smaller than the true value obtained numerically. 
We note that the computation of the Witten index here does not rely on translation invariance, and thus applies to a model with random couplings such as the supersymmetric SYK model \cite{Gross_arXiv, Maldacena_arXiv}. In fact, the authors of Ref. \cite{Maldacena_arXiv} carried out a similar analysis and obtained consistent results.

Though the Witten index gives a lower bound for $Z$, it does not tell us their exact values. Another powerful tool to study $Z$ is cohomology of $Q$~\cite{PRL_Fendley_2003, PRL_Fendley_2005}. The nontrivial cohomology classes of $Q$ are in one-to-one correspondence with the zero-energy ground states. Therefore, it would be interesting to use cohomology to compute $Z$ exactly. This is, however, beyond the scope of the present study and is left for future work. 

\bigskip

\section{Spontaneous SUSY Breaking}
\label{sec:break}

In this section, we show that spontaneous SUSY breaking occurs in the $\mathbb{Z}_2$ Nicolai model. We start with a precise definition of spontaneous SUSY breaking~\cite{U1Nicolai}. 

\bigskip

\noindent
{\bf Definition:} {\it SUSY is said to be spontaneously broken if the ground-state energy per site is strictly positive}. 

\bigskip

For finite-size systems, the definition simply states that SUSY is spontaneously broken if there is no zero-energy state. We note that the Hamiltonian is non-negative by construction. For the infinite-size system, the definition excludes the possibility that SUSY is restored in the infinite-volume limit, as pointed out by Witten \cite{Witten_NPB82}. Thus, the definition is applicable to both finite and the infinite-size systems. Below we discuss the two cases separately. 

\medskip

\subsection{Spontaneous SUSY breaking in finite chains}
\label{sec:SSB_finite}
Let us prove that SUSY is spontaneously broken in finite systems when $g>0$. The proof is parallel to the one given in \cite{U1Nicolai}. We first note that the following operator
\begin{align}
O_j =c_j^\dagger \bigg[ & 1-\frac{1}{g}(c_{j+1}c_{j+2}-c_{j-1}c_{i+1}+c_{j-2}c_{j-1}) \nonumber \\
 & +\frac{2}{g^2}c_{j-2}c_{j-1}c_{j+1}c_{j+2} \bigg]
\end{align}
satisfies $\{Q,O_j \}=g$ for all $j=1,2,...,N$. This can be verified by a straightforward calculation. Next, we suppose that there exists a zero-energy state $|\psi_0\rangle\neq0$. Then, from the fact that the state $|\psi_0\rangle$ is annihilated by both $Q$ and $Q^\dagger$, we find
\begin{align}
\langle\psi_0|\{Q,O_j \}|\psi_0\rangle=0.
\end{align}
However, this contradicts the fact that $g>0$. Thus, we have no zero-energy state unless $g=0$. This proves the spontaneous SUSY breaking. 

We note that the absence of zero-energy states simply implies $W={\rm Tr}[(-1)^F e^{-\beta H}]=0$ for $g > 0$. The discontinuity of the Witten index at $g=0$ can be understood as follows. As we see in the next section, the dispersion of excitations becomes flatter and flatter with decreasing $g$, and is completely flat in the limit $g=0$. Thus, we expect that a large number of excitation energies go to zero simultaneously when approaching $g=0$, and the abrupt change in the Witten index is allowed.

\subsection{SUSY breaking in the infinite-volume limit}
\label{sec:SSB_infinite}

Let us next prove that SUSY is spontaneously broken in the infinite-volume limit when $g$ is sufficiently large. To this end, we use Anderson's argument~\cite{PR_Anderson,PRB_Valent,PRL_Nie,Nie_arXiv,PRD_Beccaria}, which gives a lower bound for the ground-state energy.
The proof is again parallel to the one in \cite{U1Nicolai}.

Let $E_0$ be the true ground-state energy of the chain of length $N$. Because the sum of the lowest energies of the individual terms in Eq. (\ref{eq:hamz2}) is equal to or less than $E_0$, we get the following inequality:
\begin{align}
E_0\ge Ng^2+E_0^{\rm free},
\label{eq:gs_inq}
\end{align}
where $E_0^{\rm free}$ is the ground-state energy of $H_{\rm free}$. Here, we have used the fact that the ground-state energy of $H_{1}+H_{2}$ (the Hamiltonian for $g=0$) is zero, as shown in the previous section. 
Dividing both sides of Eq. (\ref{eq:gs_inq}) by $N$, we have the following inequality
\begin{align}
e(N)\ge g^2+e^{\rm free}(N),
\end{align}
where we denote the ground-state energies per site of $H$ and $H_{\rm free}$ by $e(N)$ and $e^{\rm free}(N)$, respectively. This inequality is valid for all $N$. In the infinite-volume limit, $e^{\rm free}(N)$ becomes $-4g/\pi$. A detailed derivation is given in Appendix \ref{sec:free}. From the result, we find that the ground-state energy per site in the infinite-volume limit, say $e_0$, is bounded from below as
\begin{align}
e_0\ge g\left(g-\frac{4}{\pi}\right).
\label{eq:inf_ineq}
\end{align}
Therefore, it is clear that SUSY is broken spontaneously in the infinite system when $g>4/\pi$. The condition is sufficient but may not be necessary. In fact, numerical results suggest that spontaneous SUSY breaking occurs in the infinite-volume limit unless $g=0$. We expect that a more sophisticated method can prove this rigorously, but leave this possibility for future work.

\section{Nambu-Goldstone fermions}
\label{sec:NGf}

In the previous section, we have shown that SUSY is spontaneously broken when parameter $g$ is larger than $0$ ($4/\pi$) for finite (infinite) systems. In this section, we show the existence of massless fermionic excitations. In Sec.~\ref{sec:var}, we prove, with a variational argument, the existence of an excited state whose excitation energy is bounded from above by a linear dispersion relation. In Sec.~\ref{sec:ED}, we show numerical results obtained by exact diagonalization. The results provide convincing evidence that the lowest excited states have  cubic dispersion. 

\subsection{Variational argument}
\label{sec:var}

In this subsection, we prove that spontaneous SUSY breaking leads to the existence of massless fermionic excitations. We use the Bijl-Feynman ansatz \cite{PR_Feynman}, which was used to study the low-lying excitations of the Heisenberg antiferromagnets~\cite{ZPB_Horsch, PRB_Stringari, JPSJ_Momoi}. We assume the condition that $g>4/\pi$ so that SUSY is spontaneously broken, and that the ground state degeneracy does not increase as system size $N$ increases. This is reasonable because our numerical results suggest that the ground-state degeneracy is two when $N$ is odd, while it is four when $N$ is even.

Let $|\psi_0\rangle$ be a normalized ground state of $H$. Without loss of generality, we assume that $|\psi_0\rangle$ is annihilated by the supercharge $Q$. The state $Q^\dagger |\psi_0\rangle$ is another ground state with the same energy. Since the fermionic parity $(-1)^F$ and the inversion-like operator $U$ in Eq. (\ref{eq:inversion_like}) commute with $H$, $|\psi_0\rangle$ can be chosen to be an eigenstate of $(-1)^F$ and $U$. Note that the eigenvalues of $U$ take the form $e^{i\theta}$ ($\theta \in \mathbb{R}$) because of the unitarity of $U$. For the purpose of our discussion, we define local supercharges as
\begin{align}
q_j=gc_j+c_{j-1}c_jc_{j+1}.
\end{align}
The Fourier transform of $q_j$ is then defined as
\begin{align}
Q_p:=\sum_{j=1}^Ne^{-{\mathrm i}pj}q_j.
\end{align}
Here, the wave number $p$ takes values $p=2\pi m/N$ ($m\in\mathbb{Z}$). We note that $Q_p$ becomes the supercharge $Q$ when $p=0$. The operators $Q_p$ and $Q_{-p}$ ($Q_p^\dagger$ and $Q_{-p}^\dagger$) are related to each other by $U$ as
\begin{align}
U^{-1}Q_pU=\mathrm{i}Q_{-p} \ , \ U^{-1}Q^\dagger_pU=-\mathrm{i}Q^\dagger_{-p} .
\label{UQ2}
\end{align}

We now introduce the following variational states
\begin{align}
|\psi_{1p}\rangle=(Q_p+Q_p^\dagger)|\psi_0\rangle \ , \ |\psi_{2p}\rangle=\mathrm{i}(Q_p^\dagger-Q_p)|\psi_0\rangle,
\end{align}
and assume $p \neq 0$. 
Here, the states $|\psi_{1p}\rangle$ and $|\psi_{2p}\rangle$ are orthogonal to the ground states $|\psi_0\rangle$ and $Q^\dagger|\psi_0\rangle$ since $|\psi_{ip}\rangle$ ($i=1,2$) is a linear combination of two states with nonzero momenta $+p$ and $-p$ \footnote{From numerical results, the eigenvalues of the translation operator are $\pm1$ in the ground states (see Figs. \ref{fig:dispO} and \ref{fig:dispE}), which means that trial states are orthogonal to all the ground states when $p$ is finite and small enough.}.
We get the variational energy as follows:
\begin{align}
\epsilon_{\rm var}(p) & =\frac{1}{2}\left(\frac{\langle\psi_{1p}|H|\psi_{1p}\rangle}{\langle\psi_{1p}|\psi_{1p}\rangle}+\frac{\langle\psi_{2p}|H|\psi_{2p}\rangle}{\langle\psi_{2p}|\psi_{2p}\rangle}\right)-E_0 \nonumber \\
& =\frac{\langle[Q_p,[H,Q_p^\dagger]]\rangle_0}{\langle\{Q_p^\dagger,Q_p\}\rangle_0}.
\end{align}
Here, the symbol $\langle\cdots\rangle_0$ denotes the expectation value in the ground state, i.e., $\langle\psi_0|\cdots|\psi_0\rangle$.

To evaluate the numerator, it is important to note that the local supercharges satisfy the following locality,
\begin{align}
\{q_i,q_j^\dagger\}=\left\{ \begin{array}{ll}
{\rm nonzero} & |i-j|\le2 \\
0 & {\rm otherwise}
\end{array} \right. .
\label{local2}
\end{align}
From the above relations and the identity $[H, Q_p^\dagger]=[Q^\dagger,\{Q,Q_p^\dagger\}]$, we find that the commutator $[H, Q_p^\dagger]$ is a sum of local operators. However, $[Q_p,[H,Q_p^\dagger]]$ may not be local. 
To obtain an upper bound of the dispersion, we use the Pitaevskii-Stringari inequality~\cite{JLTP_Pitaevskii},
\begin{align}
|\langle \psi | [A^\dagger, B] | \psi \rangle|^2 
\le \langle \psi | \{ A, A^\dagger \} | \psi \rangle 
\langle \psi | \{ B, B^\dagger \}| \psi \rangle,
\end{align}
which holds for any state $|\psi \rangle$ and arbitrary operators $A$, $B$. The proof of this inequality can be found in \cite{JLTP_Pitaevskii, U1Nicolai}. 
With this inequality, we have
\begin{align}
\epsilon_{\rm var}(p)^2\le\frac{\langle\{[Q_p,H],[H,Q_p^\dagger]\}\rangle_0}{\langle\{Q_p^\dagger,Q_p \}\rangle_0}.
\label{eq:evar}
\end{align}
For clarity, we introduce two functions of wave number $p$ defined by
\begin{align}
f_{\rm d}(p) & =\langle\{Q_p,Q_p^\dagger\}\rangle_0, \\
f_{\rm n}(p) & =\langle\{[Q_p, H], \, [H, Q_p^\dagger] \}\rangle_0,
\end{align}
in terms of which Eq. (\ref{eq:evar}) is rewritten as
\begin{align}
\epsilon_{\rm var}(p) \le \sqrt{\frac{f_{\rm n}(p)}{f_{\rm d}(p)}}.
\label{eq:evar2}
\end{align}

First, let us examine $f_{\rm d}(p)$. With Eq. (\ref{UQ2}) and the fact that $|\psi_0\rangle$ is an eigenstate of $U$, we find that $f_{\rm d}(p)$ is an even function of $p$. Then it follows from $f_{\rm d}(0)=E_0$ that we have
\begin{align}
f_{\rm d}(p)=N\left(\frac{E_0}{N}+O(p^2)\right).
\end{align}
Here, $f_{\rm d}(p)$ is of the order of $N$ (with the ground-state energy density $E_0/N$ fixed) from the locality Eq. (\ref{local2}) and is non-vanishing for small $p$. A more precise estimate of $f_{\rm d} (p)$ is presented in Appendix \ref{sec:fdp}. Next, we examine the numerator $f_{\rm n}(p)$. Using the locality conditions Eq. (\ref{local2}), we find that $f_{\rm n}(p)$ is also of the order of $N$. With Eq. (\ref{UQ2}) again, we find that $f_{\rm n}(p)$ is an even function of $p$. Since $Q$ and $Q^\dagger$ are conserved charges ($[H,Q]=[H,Q^\dagger]=0$), we have $f_{\rm n}(0)=0$. Therefore, we get
\begin{align}
f_{\rm n}(p)=N\left(Cp^2+O(p^4)\right),
\end{align}
where $C$ is a non-negative constant. Then we have
\begin{align}
\epsilon_{\rm var}(p)\le \sqrt{\frac{C}{E_0/N}} |p|+O(p^2).
\label{eq:ubz2}
\end{align}
This clearly shows the existence of massless excitations. Since the trial states and the ground state $|\psi_0\rangle$ have the opposite parities, these excitations can be thought of as fermionic ones,  
i.e., Nambu-Goldstone fermions.

In the above argument, the assumption that the ground state degeneracy is finite and constant for the same parity of $N$ plays an important role since the possibility that the trial states become other ground states orthogonal to $|\psi_0\rangle$ and $Q^\dagger|\psi_0\rangle$ for small $p$ can be excluded.

\subsection{Numerical result}
\label{sec:ED}
In the previous subsection, we proved the existence of massless excitation. One might think that the inequality Eq. (\ref{eq:ubz2}) implies a linear dispersion. However, the actual dispersion in the $\mathbb{Z}_2$ Nicolai model is most likely to be cubic in $p$. In order to verify this, we first consider the large-$g$ limit. In the large-$g$ limit, the Hamiltonian is dominated by the constant term and the free part,
\begin{align}
H\sim g^2N+H_{\rm free}.
\end{align}
Hamiltonians $H_{1}$ and $H_{2}$ in Eq. (\ref{eq:hamz2}) are negligible since they are independent of $g$. Using the Bogoliubov transformation, Hamiltonian $H_{\rm free}$ can be rewritten as 
\begin{align}
H_{\rm free} & =2g\sum_{0 <p \le \pi}(|f(p)|d_{1p}^\dagger d_{1p} -|f(p)|d_{2p}^\dagger d_{2p}).
\label{eq:BGtr}
\end{align}
Here, $f(p)=\mathrm{i} [2{\rm sin}(p)-{\rm sin}(2p)]$ and the momentum $p$ takes values $2\pi m/N$ ($m\in \mathbb{Z}$) . Precise definitions of the quasiparticle operators, $d_{1p}$,$d_{2p}$, can be found in Appendix \ref{sec:free}. From Eq. (\ref{eq:BGtr}), one can see that the lowest-lying excitation energy of $H_{\rm free}$ is given by $E(p)=2g|f(p)|$. When $p$ is small enough ($p\ll\pi$), $E(p)$ can be approximated as
\begin{align}
E(p)  =2g|2{\rm sin}(p)-{\rm sin}(2p)| \nonumber 
\sim 2g|p|^3.
\end{align}
Thus, the dispersion is indeed cubic in the large-$g$ limit.

\begin{figure}[h]
\includegraphics[width=1.0\columnwidth]{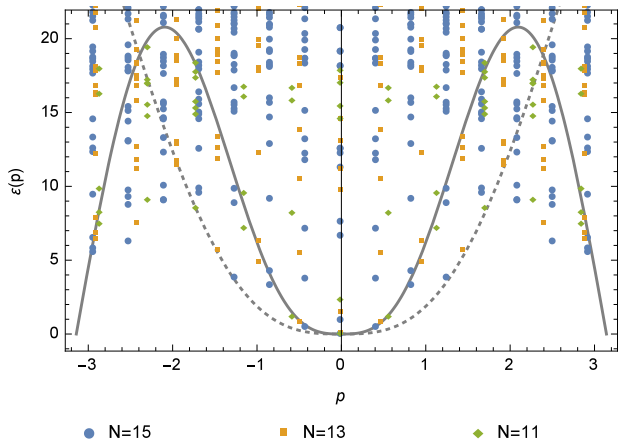}
\caption{Dispersion relation of the $\mathbb{Z}_2$ Nicolai model for $g=4$. We plot energy spectrum $\varepsilon(p)$ for $N=11,13,15$. Here, $p$ is the wave number. The gray solid curve indicates the one-particle dispersion relation of $H_{\rm free}$ and is described by $2g|f(p)|$. Gray dotted curve is described by $4g|f(p/2)|$ and indicates the dispersion of two-particle bound states of $H_{\rm free}$.}
\label{fig:dispO}
\end{figure}

\begin{figure}[h]
\includegraphics[width=1.0\columnwidth]{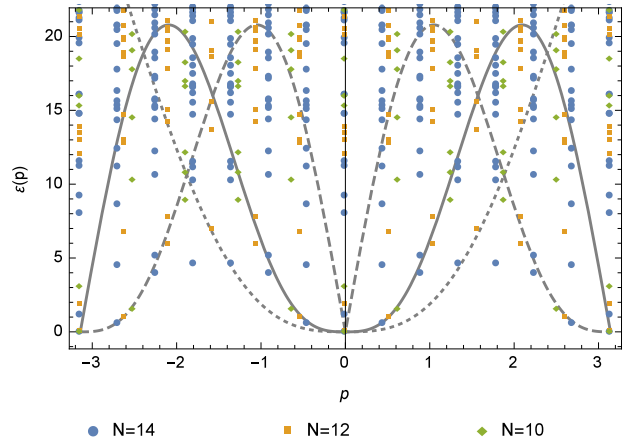}
\caption{Dispersion relation of the $\mathbb{Z}_2$ Nicolai model for $g=4$. We plot energy spectrum $\varepsilon(p)$ for $N=10,12,14$. Here, $p$ is the wave number. The gray solid curve indicates the one-particle dispersion relation of $H_{\rm free}$ and is described by $2g|f(p)|$. The gray dotted and dashed curves are described by $4g|f(p/2)|$ and $2g|f(\pi-p)|$, respectively. They indicate the dispersion of two-particle bound states of $H_{\rm free}$.}
\label{fig:dispE}
\end{figure}

Now the question is whether or not the dispersion is cubic for moderate values of $g$. To see this, we calculated dispersion relation with exact diagonalization with $N=10,\dots,15$. The energy spectrum of the total Hamiltonian with PBC for $g=4$ is shown in Fig. \ref{fig:dispO} and Fig. \ref{fig:dispE}. In Fig. \ref{fig:dispO}, we display energy spectra for odd $N$, while we display those for even $N$ in Fig. \ref{fig:dispE}.

In Fig.~\ref{fig:dispO}, the gray solid curve is the one-particle excitation spectrum of the free Hamiltonian $H_{\rm free}$ for $g=4$ as a function of $p$, and the gray dotted curve is the two-particle spectrum with total momentum $P$, which is described as $4g|f(P/2)|$. The dispersion fits to the gray solid curve and is quite likely to be cubic in the vicinity of $p=0$. We expect that the energy levels below the gray dotted curve correspond to those of $m$-particle bound states with $m >2$. 
It is known in ferromagnetic spin chains that some energy eigenvalues of bound states are lower than those of scattering states~\cite{Haldane_1982}.

The results of exact diagonalization for even $N$ with PBC are shown in Fig.~\ref{fig:dispE}. The definitions of gray solid and dotted curves are the same as those in Fig.~\ref{fig:dispO}. In Fig.~\ref{fig:dispE}, the dispersion relation is again likely to be cubic around $p=0$ since it fits the gray solid curve. From the plot, we see that the energy spectrum is symmetric about $p=\pi/2$. Because of this symmetry, we have also cubic dispersion around $p=\pi$. The gray dashed curve is described by $2g|f(\pi-p)|$ which shows good agreement with the data.

\begin{figure}[h]
\includegraphics[width=0.93\columnwidth]{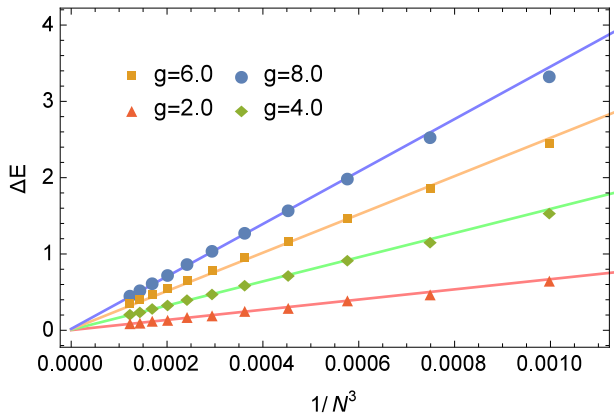}
\caption{Energy difference between the first excited and the ground states as a function of $1/N^3$ for $g=2,4,6,8$. 
Lines are fits to the data of $N=15,\dots,20$.}
\label{fig:1st}
\end{figure}

In order to provide further evidence of cubic dispersion, we plot the first excitation energies relative to the ground state as a function of $1/N^3$ for $g=2,4,6,8$. They are calculated using exact diagonalization with $N=10,\dots, 20$. The results are shown in Fig. \ref{fig:1st}.
Since $1/N$ is proportional to the wave number $p$, the Fig. \ref{fig:1st} tells us that the lowest excitation energy is cubic in $p$. This result is consistent with the excitation spectrum shown in Figs. \ref{fig:dispO} and \ref{fig:dispE} and the single-particle energy spectrum of $H_{\rm free}$ in Eq. (\ref{eq:BGtr}). Therefore, the NG fermions in our system have cubic dispersion at low energies.  

An interesting question is whether this cubic dispersion is stable to the addition of small perturbations. We partially answer this question by using a scaling argument and numerical calculations. We find that the dispersion in the vicinity of $p=0$ is still cubic even in the presence of SUSY-preserving perturbations (see Appendix~\ref{sec:ModifyQ} for details). 

\section{Conclusion}
\label{sec:conclusion}
In this paper, we have introduced and studied a lattice fermion model without ${\rm U}(1)$ symmetry but with $\mathbb{Z}_2$ symmetry in one dimension, whose Hamiltonian is defined as the anticommutator of the supercharges $Q$ and $Q^\dagger$. 
When $g=0$, SUSY is unbroken and the ground-state entropy is extensive, i.e., the number of the zero-energy states grows exponentially with the system size $N$. 
When $g$ is nonzero, SUSY is spontaneously broken in finite chains. For the infinite chain, we showed that SUSY is spontaneously broken when $|g|>4/\pi$. We numerically found that the number of the ground states is finite and depends only on the parity of $N$ unless $g=0$. 
F

Our analysis has revealed the nature of the low-lying excitations in the $\mathbb{Z}_2$ Nicolai model. With a variational approach, we proved that there exist low-lying states whose energies are bounded from above by a linear dispersion. This result clearly shows the existence of gapless excitations. Furthermore, using exact diagonalization, we showed that the dispersion relation of our model is cubic in wave number $p$, $\omega\propto |p|^3$, at low energies.

The low-lying excitation of the extended Nicolai model, which has ${\rm U}(1)$ symmetry, is described by a conformal field theory with central charge $c=1$ \cite{U1Nicolai}. Thus, this low-energy effective theory has emergent Lorentz invariance. By contrast, the low-energy theory of the model introduced in this paper explicitly breaks Lorentz invariance since the dispersion is cubic.

The model we introduced is the first example in which NG fermions with cubic dispersion are realized by the spontaneous SUSY breaking. This dispersion relation is protected unless one introduces extra degrees of freedom, and one cannot lower its power as long as we preserve the SUSY. 
This is remarkable for the following two reasons. First of all, from the algebraic structure and the symmetry breaking pattern alone, we may expect the linear dispersion relation (e.g. as in the ${\rm U}(1)$ symmetric generalized Nicolai model \cite{U1Nicolai}), but we encounter the cubic dispersion. Secondly, we, nevertheless cannot modify the dispersion relation, and the cubic dispersion relation is actually stable.

Gapless excitations are at the core of universality and non-trivial infrared physics. One may use our exotic cubic dispersion relation with non-trivial dynamical critical exponent to explore new phases of condensed matters. In this respect, we note that our model can be extended to higher dimensions. In Appendix \ref{sec:2dSUSY}, we demonstrate that a generalization of our model to the two-dimensional triangular lattice also exhibits NG fermions with cubic dispersion in the large-$g$ limit. In particular, we find that the two-dimensional dispersion relation of our NG fermions is the same as that of the Majorana fermion excitations discussed in a class of quantum spin liquids on the triangular lattice~\cite{PRB_Biswas,PRB_Wang}. Given the same dispersion relation, it is an interesting question to ask if we can realize SUSY and its spontaneous breaking in such quantum spin systems.

\bigskip

\begin{acknowledgments}
The authors thank Hajime Moriya and Yutaka Akagi for valuable discussions. H. K. was supported in part by JSPS KAKENHI Grants No. JP15K17719 and No. JP16H00985.
Y. N. is supported in part by the Rikkyo University Special
Fund for Research and MEXT-Supported Program for the Strategic Research Foundation at Private Universities, 2014-2017.

\end{acknowledgments}

\appendix
\section{Zero Energy States in $g=0$ open chains}
\label{sec:openZES}
Here we present our results for the number of ground states in the model with $g=0$ and open boundary conditions. The supercharge for the open chain takes the following form:
\begin{equation}
Q=\sum_{j=1}^{N-2} c_jc_{j+1}c_{j+2}.
\end{equation}
As in the periodic case, SUSY is unbroken in this case. We numerically calculated the number of the zero-energy grounds states $Z$ for chains of length $N=3, 4, ..., 16$. The results obtained are summarized in the second row of Table \ref{tab:OZES}. 

\begin{widetext}
\begin{center}
\begingroup
\renewcommand{\arraystretch}{1.3}
\begin{table}[htb]
\begin{tabular}{ccccccccccccccc}
\hline \hline
~~~~~$N$~~~~~ & \,~3~\, & 4 & 5 &6 & 7 & 8 & 9 & 10 & 11 & 12 & 13 & 14 & 15 & 16  \\ \hline
~~~~~$Z$~~~~~ & ~6 ~ & ~12~ & ~20 ~ & ~36~ & ~64 ~ & ~112~ & ~200~ & ~352~ & ~624~ & ~1104~ & ~1952~ & ~3456~ & ~6112~ & ~10816~  \\
~~~~$Z_{\rm cl}$~~~~ & \,~6~\, & ~10~ & ~16~ & ~26~ & ~42~ & ~68~ & ~110~ & ~178~ & ~288~ & ~466~ & ~754~ & ~1220~ & ~1974~ & ~3194~ 
\\
~~~~~$W$~~~~~ & \,~6~\, & ~12~ & ~18~ & ~36~ & ~54~ & ~108~ & ~162~ & ~324~ & ~486~ & ~972~ & ~1458~ & ~2916~ & ~4374~ & ~8748~
\\
\hline \hline
\end{tabular}
\caption{Ground-state degeneracy of the open chain with $g=0$ up to $N=16$ sites. $Z$, $Z_{\rm cl}$, and $W$ refer to the number of zero-energy states, the number of classical ground states, and the Witten index, respectively.}
\label{tab:OZES}
\end{table}
\endgroup
\end{center}
\end{widetext}

We find that the numbers $Z$ follow the sequence A107383 in the On-Line Encyclopedia of Integer Sequences \footnote{The On-Line Encyclopedia of Integer Sequences, http://oeis.org}, which is defined by the following recurrence relation,
\begin{align}
Z_N=2 Z_{N-2}+2 Z_{N-3}, \quad \ Z_{0}=1,~ Z_{1}=2,~ Z_{2}=4,
\end{align}
where we denote by $Z_N$ the number of the ground states of the chain of length $N$. For large $N$, this number scales as $Z_N \sim (x^*)^N$, where $x^* \sim 1.769$ is the real root of the cubic equation $x^3-2x-2=0$. Although we cannot prove this analytically, we believe that the result holds for arbitrary $N$. 

We remark that the result extends to the inhomogeneous case where the supercharge takes the form
\begin{equation}
Q=\sum_{j=1}^{N-2} s_j\, c_jc_{j+1}c_{j+2},
\end{equation}
with spatially varying couplings $s_j \ne 0$ ($j=1,2,...,N$). Surprisingly, our numerical results suggest that the number of the ground states remains unchanged for an arbitrary set of $s_j$. Therefore, we conjecture that the numbers $Z_N$ are robust against perturbations that make $Q$ inhomogeneous.

The number of the classical ground states ($Z_{\rm cl}$) can be computed in the same fashion as in the periodic case discussed in the main text. 
The analytic expression for $Z_{\rm cl}$ is given by
\begin{align}
Z_{\rm cl} = \frac{2}{5}
\Bigg[
  &(5-2\sqrt{5}) \left( \frac{1-\sqrt{5}}{2} \right)^{N-2} \nonumber \\
&+(5+2\sqrt{5}) \left( \frac{1+\sqrt{5}}{2} \right)^{N-2}
\Bigg].
\end{align}
Their numerical values are shown in the third row of Table \ref{tab:OZES}, along with the Witten indices in the forth row. Clearly, they are equal to or smaller than the exact values of $Z$.

\section{SUSY SYK model}
\label{sec:SYK}
The  $\mathbb{Z}_2$ Nicolai model discussed in the main text was defined on a one-dimensional lattice. Instead, let us consider the infinite-range random supercharge 
\begin{align}
Q = \sum^N_{i=1} g_i c_i + \sum_{i,j,k=1}^N C_{ijk} c_i c_j c_k \ , \label{SUSYSYK}
\end{align} 
where $g_i$ and $C_{ijk}$ are random Gaussian variables with the variance
\begin{align}
\langle g_i g_i^* \rangle = \frac{2g}{N^2}, \quad
\langle C_{ijk} {C}^*_{ijk} \rangle = \frac{2J}{N^2} \ .
\end{align}

Let us first consider the case with $g_i=0$ ($i=1,...,N$). 
Defining the Hamiltonian by $H = \{Q,Q^\dagger\}$, we obtain the SUSY version of the SYK model, which is also discussed in \cite{Gross_arXiv, Maldacena_arXiv}. The original SYK model has an infinite-range random four-Fermi interaction and possesses a non-trivial large-$N$ solution. It is supposed to describe a holographic dual of a black hole. In a similar manner, we expect that the SUSY version of the SYK model with the supercharge \eqref{SUSYSYK} describes a holographic dual of a supersymmetric or extremal black hole (effectively in $1+1$ space-time dimensions).

Some of the interesting features of the $\mathbb{Z}_2$ Nicolai model studied in the main text are shared with the SUSY version of the SYK model. For example, the ground state of the model for fixed $C_{ijk}$ is exponentially degenerate. The number of the ground states and the Witten index for various $N$ are shown in Table \ref{tab:SYK}. In most cases, the inequality $Z \ge W$ is saturated. The exceptional cases where $Z > W$ are consistent with those found in Ref. \cite{Maldacena_arXiv}.

\begin{widetext}
\begin{center}
\begingroup
\renewcommand{\arraystretch}{1.3}
\begin{table}[htb]
\begin{tabular}{ccccccccccccccc}
\hline \hline
~~~~~$N$~~~~~ & \,~3~\, & 4 & 5 &6 & 7 & 8 & 9 & 10 & 11 & 12 & 13 & 14 & 15 & 16  \\ \hline
~~~~~$Z$~~~~~ & \,~6~\, & ~12~ & ~20~ & ~36~ & ~54 ~ & ~108~ & ~168~ & ~324~ & ~486~ & ~972~ & ~1460~ & ~2916~ & ~4374~ & ~8748~  \\
~~~~~$W$~~~~~ & \,~6~\, & ~12~ & ~18~ & ~36~ & ~54~ & ~108~ & ~162~ & ~324~ & ~486~ & ~972~ & ~1458~ & ~2916~ & ~4374~ & ~8748~
\\
\hline \hline
\end{tabular}
\caption{Ground-state degeneracy of the SUSY SYK model with $g_i=0$ and fixed $C_{ijk}$ up to $N=16$ sites. $Z$ and $W$ refer to the number of zero-energy states and the Witten index, respectively.}
\label{tab:SYK}
\end{table}
\endgroup
\end{center}
\end{widetext}

One may formally study the large-$N$ scaling solution for the (quenched-average) two-point function
\begin{align}
\langle c^\dagger(\tau) c(0) \rangle = C \frac{\mathrm{sgn}(\tau)}{|J\tau|^{1/3}} \ 
\end{align}
by solving the Schwinger-Dyson equation with the scaling ansatz \footnote{See also \cite{Gross_arXiv}. Our result was communicated to V.~Rosenhaus by one of the authors (Y.~N.) at Strings 2016.}. We, however, note that the model has exponentially degenerate ground states, so the meaning of the scaling solution should be understood better. 

As in the $\mathbb{Z}_2$ Nicolai model discussed in the main text, having non-zero $g$ makes the  SUSY spontaneously broken (for a fixed set of $g_i$ and $C_{ijk}$). One would expect a Nambu-Goldstone mode, but since the model is not translationally invariant, it is more non-trivial to discuss the dispersion relation. 
It is an interesting future direction to see if such a deformation is related to black holes with spontaneous SUSY breaking.

\section{Ground-state energy of auxiliary free-fermion problem}
\label{sec:free}
In this appendix, we calculate the exact ground-state energy of the free Hamiltonian Eq. (\ref{eq:freeham}), with PBC. 
The Hamiltonian in Fourier space is
\begin{align}
H_{\rm free} = 2g\sum_{0 <p \le \pi} (c^\dagger_p,c_{-p})
\begin{pmatrix}
0 & f^\ast(p) \\
f(p) & 0
\end{pmatrix}
\begin{pmatrix}
c_p \\
c^\dagger_{-p}
\end{pmatrix}
\nonumber \\ \quad {\rm with} \quad f(p) =\mathrm{i}({\rm sin}(2p) -2{\rm sin}(p)),
\end{align}
where $c_p := \sum_j e^{-\ii pj}c_j/\sqrt{N}$ and the wave number $p$ takes the values
\begin{equation}
p = \frac{2\pi}{N} \ell, \quad \quad \ell \in \mathbb{N}. 
\end{equation}
After the Bogoliubov transformation, the Hamiltonian reads,
\begin{align}
H_{\rm free}=2g\sum_{0 <p \le \pi}(|f(p)|d_{1,p}^\dagger d_{1,p}-|f(p)|d_{2,p}^\dagger d_{2,p}).
\end{align}
Here, $d_{1,p}$ and $d_{2,p}$ are quasiparticle operators which are defined as $d_{1,p}:=(c_p^\dagger-\mathrm{i}c_{-p})/\sqrt{2}$, and $d_{2,p}:=(c_p^\dagger+\mathrm{i}c_{-p})/\sqrt{2}$, respectively. They satisfy ordinary anticommutation relations,
\begin{align}
\{d_{i,p},d_{j,p'}\}=0, \quad \{d_{i,p},d^\dagger_{j,p'}\}=\delta_{i,j}\delta_{p,p'} \quad (i,j=1,2).
\end{align}
One of the ground states of $H_{\rm free}$ is fully filled by quasiparticles $d_{2,p}$ with negative energies, i.e.,
\begin{align}
\prod_{0<p \le \pi}d_{2,p}^\dagger |{\rm vac}\rangle,
\end{align}
where $|{\rm vac}\rangle$ is the vacuum ($d_{2,p}|{\rm vac}\rangle=d_{1,p}|{\rm vac}\rangle=0$ for all $p$). One finds the ground-state energy of $H_{\rm free}$ as
\begin{align}
E_0^{\rm free}=-2g\sum_{0<p\le\pi}|f(p)|.
\end{align}
Since $f(\pi)=0$, we can add $p=\pi$ to the sum. For a chain of even length $N$, one finds
\begin{align}
E_0^{\rm free}=-\frac{4g}{{\rm tan}(\pi/N)},
\end{align}
while, for odd $N$, one gets
\begin{align}
E_0^{\rm free}=-g\frac{2({\rm cos}(\pi/N)+1)^2}{{\rm sin}(2\pi/N)}.
\end{align}
Both the ground-state energies approach the same value $E_0^{\rm free} \sim -4gN/\pi$ in the infinite-volume limit $N\to \infty$.

\section{A lower bound for $f_{\rm d}(p)$}
\label{sec:fdp}
Similarly to the lower bound for the ground-state energy, we have
\begin{equation}
\{ Q_p, Q^\dagger_p  \} \ge H_{\rm free} (p) + g^2N.
\label{eq:Hamp}
\end{equation}
Here, we write $A \ge B$ to denote that $A-B$ is positive semidefinite. 
The modified free Hamiltonian $H_{\rm free} (p)$ is defined as
\begin{align}
H_{\rm free} (p) & := 2g \cos p\, \sum^N_{j=1} (c_{j+1}^\dagger c_j^\dagger+c_jc_{j+1})\nonumber \\
& -g\sum_{j=1}^N(c_{j+1}^\dagger c_{j-1}^\dagger +c_{j-1}c_{j+1}).
\end{align}
As in the case of $H_{\rm free}$ discussed in Appendix \ref{sec:free}, whether the length of the chain is even or odd is important in the calculation of the ground-state energy of a free hopping Hamiltonian \cite{PRL_Nie,U1Nicolai}.  

Let us derive the condition under which the denominator of Eq. (\ref{eq:evar2}) becomes positive. First, we examine the case of even $N$. A straightforward calculation shows that the operator $H_{\rm free}$ is bounded from below as
\begin{align}
H_{\rm free}(p)\ge -\frac{4g \cos p}{{\tan}(\pi/N)}.
\end{align}
Applying this inequality to Eq. (\ref{eq:Hamp}), we get
\begin{align}
\{Q_p,Q_p^\dagger \}\ge Ng\left(g-\frac{4}{\pi} \cos p\right).
\end{align}
When $g>(4/\pi) \cos p$, the dominator of Eq. (\ref{eq:evar2}) must be positive. Since $\cos p$ is equal to or smaller than unity, the dominator of Eq. (\ref{eq:evar2}) must always be positive when $g>4/\pi$, in which case SUSY is spontaneously broken.

Next, we examine the chain of odd length $N$. When $N$ is odd, we get the following inequality
\begin{align}
H_{\rm free}(p)\ge -2g\left(\frac{{\cos}(\frac{\pi}{N})+1}{{\sin}(\frac{\pi}{N})} \cos p+\frac{{\sin}^2(\frac{\pi}{N})}{{\sin}(\frac{2\pi}{N})}\right).
\end{align}
This inequality is more complex than that for even $N$. 
For $|p|< \frac{\pi}{3}$ and $N \ge 4$, the right-hand side of this inequality is bounded from below as follows:
\begin{align}
({\rm RHS}) &= -2 g \frac{1+ \cos (\frac{\pi}{N}) }{\sin ( \frac{2\pi}{N} )}
\left[ 1+(2 \cos p -1) \cos \left( \frac{\pi}{N} \right) \right] \nonumber \\
& \ge 2g \frac{1+ \cos (\frac{\pi}{N}) }{\sin ( \frac{2\pi}{N} )} \times 2 \cos p \nonumber \\
& \ge 2 g N \cos p,
\label{eq:LHS1}
\end{align}
where we have used the fact that $\cos p > 1/2$ and the inequality $\sin x \ge 2x \ge/\pi$ which is valid when $0 \le x \le \pi/2$. Applying the inequality Eq. (\ref{eq:LHS1}) to Eq. (\ref{eq:Hamp}), we have 
\begin{align}
\{Q_p,Q_p^\dagger\} \ge N g (g- 2 \cos p),
\end{align}
In this way, we get a rigorous lower bound for $f_{\rm d}(p)$ in Eq. (\ref{eq:evar2}). Since $\cos p$ is smaller than unity, $f_{\rm d}(p)$ must be nonvanishing for $g>2$. This gives a sufficient condition under which the inequality Eq. (\ref{eq:ubz2}) holds. 

\section{Stability of cubic dispersion}
\label{sec:ModifyQ}
In this section, we discuss the stability of the cubic dispersion of the $\mathbb{Z}_2$ Nicolai model near $p=0$ against perturbations using a scaling argument used in~\cite{PRB_Biswas} and numerical calculations. We restrict ourselves to perturbations that do not break SUSY explicitly. This can be done by adding to the supercharge $Q$ an odd polynomial in $c_j$ ($j=1,2,...,N$) which is local in space. In the continuum limit, the unperturbed supercharge 
Eq. (\ref{eq:SuQ}) is written as follows,
\begin{align}
Q=\int  dx \ (g\psi +\psi \partial\psi\partial^2\psi).
\label{eq:contQ}
\end{align}
Here, $\psi$ is a fermionic field whose scaling dimension is $1/2$ in $1+1$ dimensions, and the symbol $\partial$ denotes the spatial derivative. 
Perturbations to $Q$ in the continuum limit are written as $\psi\partial\psi\partial^3\psi$ and so on. Since the term $\psi\partial\psi\partial^3\psi$ leads to the most relevant perturbations to the Hamiltonian, below we keep only this term and neglect other terms containing higher derivatives. 
The modified supercharge in the continuum limit reads 
\begin{align}
Q_{\rm m}=\int  dx \ (g\,\psi +\psi \partial\psi\partial^2\psi+\lambda\, \psi\partial\psi\partial^3\psi),
\end{align}
from which the Hamiltonian is defined as $H_{\rm m}=\{Q_{\rm m}^\dagger, Q_{\rm m}\}$. In a Lagrangian formulation, 
perturbations in the action are written as
\begin{align}
\int dxd\tau \ \partial\psi\partial^3\psi+{\rm H.c.},
\label{eq:int1}
\end{align}
\begin{align}
\int dxd\tau \ \partial\psi^\dagger\partial^2\psi^\dagger\partial\psi\partial^3\psi+{\rm H.c.},
\label{eq:int2}
\end{align}
and so on. Here, $\tau$ is the imaginary time. From the following scaling transformation,
\begin{align}
x'=bx\quad , \quad \tau'=b^z\tau \quad {\rm and} \quad \psi'=b^{-1/2}\psi \quad (b>1),
\end{align}
we see that a quadratic term like Eq. (\ref{eq:int1}) with $s$-spatial derivatives has scaling dimension $s-z$, and a quartic term like Eq. (\ref{eq:int2}) with $s$-spatial derivatives has scaling dimension $s-z+1$, where $z$ is the dynamical critical exponent. 
From the results in the main text it is natural to assume $z=3$, in which case 
the scaling dimensions of the terms Eq. (\ref{eq:int1}) and Eq. (\ref{eq:int2}) are $1$ and $5$, respectively. 
Therefore, these terms are irrelevant and do not affect the dispersion relation. 
Other interaction terms in $H_{\rm m}$ are more irrelevant since they have higher derivatives. We emphasize that the application of the above argument is limited to perturbations that do not break SUSY explicitly. 

\begin{figure}[h]
\includegraphics[width=0.93\columnwidth]{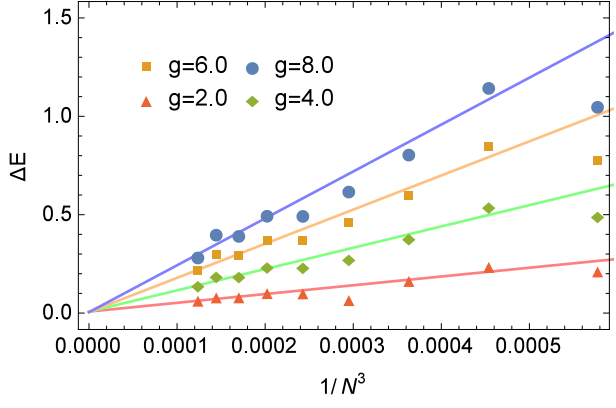}
\caption{The lowest excitation energy $\Delta E$ of $H$ [Eq. (\ref{eq:quartHam})] as a function of $1/N^3$ for $g=2,4,6,8$, $g_3=1/3$ and $g_5=1/5$. Lines are fits to the data of $N=17,\dots,20$.}
\label{fig:quintic}
\end{figure}

In order to provide supporting evidence for the above argument, we carry out numerical calculation and determine the dispersion relation. We consider the following supercharge,
\begin{align}
Q=\sum_{j=1}^N\left(gc_j+g_3c_{j-1}c_jc_{j+1}+g_5c_{j-2}c_{j-1}c_jc_{j+1}c_{j+2}\right),
\label{eq:modQ}
\end{align}
where $g_3$ and $g_5$ are parameters. The sum of the first and the second terms is identical to the supercharge of the $\mathbb{Z}_2$ Nicolai model, with the exception of the coefficient $g_3$. The Hamiltonian of this model is defined by
\begin{align}
H=\{Q^\dagger, Q\}.
\label{eq:quartHam}
\end{align}
We study the dispersion relation of the model 
using the exact diagonalization method for $N=12,\dots,20$ with periodic boundary conditions. In Fig.~\ref{fig:quintic}, we plot the first excitation energies relative to the ground state as a function of $1/N^3$ for $g=2,4,6,8$, $g_3=1/3$ and $g_5=1/5$.
The results obtained show that the first energy excitation scales with $1/N^3$ and becomes zero in the infinite-size limit. This means that the dispersion is gapless and cubic at low energies. 

From the above argument, we conclude that the cubic dispersion of the $\mathbb{Z}_2$ Nicolai model is stable against perturbations which do not break SUSY explicitly. 

\section{Generalization to two dimension}
\label{sec:2dSUSY}
In this section, we consider an extension of the $\mathbb{Z}_2$ Nicolai model to those in two dimensions. In particular, we introduce a model on a two-dimensional triangular lattice. 
We note in passing that a supersymmetric lattice fermion model on the triangular lattice has been discussed by Huijse and her collaborators \cite{NJP_Huijse}. Their model is completely different from ours since the fermion number of Huijse's model is conserved while the model discussed in this appendix does not conserve it. 

The supercharge of our model is defined as follows:
\begin{align}
Q=g\sum_{\bm{r}} c({\bm r})+\cfrac13\sum_{\bm{r}}c({\bm r})c({\bm r}+{\bm \delta}_1)c({\bm r}-{\bm \delta}_3),
\end{align}
where $\bm{r}$ denotes the position of a lattice site in the triangular lattice, $c ({\bm{r}})$ is the annihilation operator of the fermion at site ${\bm{r}}$, and ${\bm{\delta}}_i$ ($i=1,2,3$) are
the vectors along nearest neighbor bonds which satisfy the following relation,
\begin{align}
{\bm{\delta}}_1+{\bm{\delta}}_2+{\bm{\delta}}_3=0.
\end{align}
 As in the one-dimensional case, annihilation and creation operators of fermions satisfy 
\begin{align}
\{c({\bm r}),c({\bm r'})\}\!=\!\{c^\dagger({\bm r}),c^\dagger({\bm r'})\}\!=0 \ , \ \{c({\bm r}),c^\dagger({\bm r'}) \}\!=\!\delta_{{\bm r},{\bm r'}}.
\end{align}
We note that the supercharge is nilpotent, i.e., $Q^2=0$.
In the large-$g$ limit, by neglecting the four-Fermi interactions, the Hamiltonian becomes
\begin{align}
H & =\{Q,Q^\dagger\} \sim  g^2 N \nonumber \\
& +\frac{g}{3} \sum_{{\bm r},{\bm r'}}(\{c({\bm r}), c^\dagger({\bm r'})c^\dagger({\bm r'}+{\bm \delta}_1)c^\dagger({\bm r'}-{\bm \delta}_3)\}+{\rm H.c.}),
\label{eq:2dHam}
\end{align}
where $N$ is the total number of sites. The second term in Eq. (\ref{eq:2dHam}) can be written as,
\begin{align}
g \sum_{i=1,2,3} \sum_{{\bm r}} (c^\dagger ({\bm r}) c^\dagger ({\bm r}+{\bm \delta}_i) +{\rm H.c.}). 
\end{align}
Using Fourier transformation, it can be rewritten as
\begin{align}
g \sum_{i=1,2,3} \sum_{{\bm q}} ( e^{\mathrm{i}{\bm q}\cdot{\bm \delta}_i}c^\dagger({\bm q})c^\dagger(-{\bm q}) +{\rm H.c.}),
\end{align}
where $c^\dagger ({\bm q})$ denotes the Fourier transform of $c^\dagger ({\bm r})$,
and the sum runs over all momenta ${\bm q}$ included in the Brillouin zone (BZ) of the triangular lattice. By a straightforward calculation, we get the following result for any $i=1,2,3$,
\begin{align}
& g\sum_{{\bm q}}e^{\mathrm{i}{\bm q}\cdot{\bm \delta}_i}c^\dagger({\bm q})c^\dagger(-{\bm q}) \nonumber \\
= & \frac{g}{2}\sum_{\bm q}e^{\mathrm{i}{\bm q}\cdot{\bm \delta}_i}c^\dagger({\bm q})c^\dagger(-{\bm q})+\frac{g}{2}\sum_{\bm q}e^{-\mathrm{i}{\bm q}\cdot{\bm \delta}_i}c^\dagger({\bm q})c^\dagger(-{\bm q}) \nonumber \\
= & \frac{g}{2}\sum_{\bm q}(e^{\mathrm{i}{\bm q}\cdot{\bm \delta}_i}-e^{-\mathrm{i}{\bm q}\cdot{\bm \delta}_i})c^\dagger({\bm q})c^\dagger(-{\bm q}) \nonumber \\
= & \mathrm{i}g\sum_{\bm q}{\rm sin}({\bm q}\cdot{\bm \delta}_i)c^\dagger({\bm q})c^\dagger(-{\bm q}).
\end{align}
By using this result, we have
\begin{align}
H\sim g^2 N \!+\! \sum_{\vec{q}} (c^\dagger({\bm q}),c(-{\bm q}))\!
\begin{pmatrix}
0 & h^\ast({\bm q}) \\
h({\bm q}) & 0
\end{pmatrix}\!\!
\begin{pmatrix}
c({\bm q}) \\
c^\dagger(-{\bm q})
\end{pmatrix},
\end{align}
where the function $h({\bm q})$ is defined as,
\begin{align}
h({\bm q}):=\mathrm{i}g\sum_i \sin ({\bm q}\cdot{\bm \delta}_i).
\end{align}
In a two-dimensional triangular lattice, 
the three unit vectors are defined as follows: \mbox{${\bm \delta}_1=(1,0)$}, \mbox{${\bm \delta}_2=(-1/2,\sqrt{3}/2)$}, \mbox{${\bm \delta}_3=(-1/2,-\sqrt{3}/2)$}. 
With this convention, the function $h({\bm q})$ can be written in the following form:
\begin{align}
h({\bm q})\!=\!\mathrm{i}g\!\left(\! \sin (q_x)\!+\! \sin (-\frac{1}{2}q_x\!+\! \frac{\sqrt{3}}{2}q_y)\!-\! \sin (\frac{1}{2}q_x+\! \frac{\sqrt{3}}{2}q_y)\!\right)\!.
\end{align}
Here, we use ${\bm q}=(q_x,q_y)$. The one-particle energy dispersion is written as
\begin{align}
E({\bm q})=|h({\bm q})|.
\end{align}
This is identical to the dispersion of Majorana fermion excitations in a model of quantum spin liquids on a two-dimensional triangular lattice \cite{PRB_Biswas}.

\nocite{*}

\bibliography{Z2Nicolai}

\begin{thebibliography}{49}%
\makeatletter
\providecommand \@ifxundefined [1]{%
 \@ifx{#1\undefined}
}%
\providecommand \@ifnum [1]{%
 \ifnum #1\expandafter \@firstoftwo
 \else \expandafter \@secondoftwo
 \fi
}%
\providecommand \@ifx [1]{%
 \ifx #1\expandafter \@firstoftwo
 \else \expandafter \@secondoftwo
 \fi
}%
\providecommand \natexlab [1]{#1}%
\providecommand \enquote  [1]{``#1''}%
\providecommand \bibnamefont  [1]{#1}%
\providecommand \bibfnamefont [1]{#1}%
\providecommand \citenamefont [1]{#1}%
\providecommand \href@noop [0]{\@secondoftwo}%
\providecommand \href [0]{\begingroup \@sanitize@url \@href}%
\providecommand \@href[1]{\@@startlink{#1}\@@href}%
\providecommand \@@href[1]{\endgroup#1\@@endlink}%
\providecommand \@sanitize@url [0]{\catcode `\\12\catcode `\$12\catcode
  `\&12\catcode `\#12\catcode `\^12\catcode `\_12\catcode `\%12\relax}%
\providecommand \@@startlink[1]{}%
\providecommand \@@endlink[0]{}%
\providecommand \url  [0]{\begingroup\@sanitize@url \@url }%
\providecommand \@url [1]{\endgroup\@href {#1}{\urlprefix }}%
\providecommand \urlprefix  [0]{URL }%
\providecommand \Eprint [0]{\href }%
\providecommand \doibase [0]{http://dx.doi.org/}%
\providecommand \selectlanguage [0]{\@gobble}%
\providecommand \bibinfo  [0]{\@secondoftwo}%
\providecommand \bibfield  [0]{\@secondoftwo}%
\providecommand \translation [1]{[#1]}%
\providecommand \BibitemOpen [0]{}%
\providecommand \bibitemStop [0]{}%
\providecommand \bibitemNoStop [0]{.\EOS\space}%
\providecommand \EOS [0]{\spacefactor3000\relax}%
\providecommand \BibitemShut  [1]{\csname bibitem#1\endcsname}%
\let\auto@bib@innerbib\@empty
\bibitem [{\citenamefont {Wess}\ and\ \citenamefont
  {Zumino}(1974)}]{Wess_NPB74}%
  \BibitemOpen
  \bibfield  {author} {\bibinfo {author} {\bibfnamefont {J.}~\bibnamefont
  {Wess}}\ and\ \bibinfo {author} {\bibfnamefont {B.}~\bibnamefont {Zumino}},\
  }\href {\doibase http://dx.doi.org/10.1016/0550-3213(74)90355-1} {\bibfield
  {journal} {\bibinfo  {journal} {Nucl. Phys. B}\ }\textbf {\bibinfo {volume}
  {70}},\ \bibinfo {pages} {39 } (\bibinfo {year} {1974})}\BibitemShut
  {NoStop}%
\bibitem [{\citenamefont {Witten}(1982)}]{Witten_NPB82}%
  \BibitemOpen
  \bibfield  {author} {\bibinfo {author} {\bibfnamefont {E.}~\bibnamefont
  {Witten}},\ }\href {\doibase http://dx.doi.org/10.1016/0550-3213(82)90071-2}
  {\bibfield  {journal} {\bibinfo  {journal} {Nucl. Phys. B}\ }\textbf
  {\bibinfo {volume} {202}},\ \bibinfo {pages} {253 } (\bibinfo {year}
  {1982})}\BibitemShut {NoStop}%
\bibitem [{\citenamefont {Salam}\ and\ \citenamefont
  {Strathdee}(1974)}]{Salam:1974zb}%
  \BibitemOpen
  \bibfield  {author} {\bibinfo {author} {\bibfnamefont {A.}~\bibnamefont
  {Salam}}\ and\ \bibinfo {author} {\bibfnamefont {J.}~\bibnamefont
  {Strathdee}},\ }\href@noop {} {\bibfield  {journal} {\bibinfo  {journal}
  {Phys. Lett. B}\ }\textbf {\bibinfo {volume} {49}},\ \bibinfo {pages} {465}
  (\bibinfo {year} {1974})}\BibitemShut {NoStop}%
\bibitem [{\citenamefont {Volkov}\ and\ \citenamefont
  {Akulov}(1973)}]{PLB_Volkov}%
  \BibitemOpen
  \bibfield  {author} {\bibinfo {author} {\bibfnamefont {D.}~\bibnamefont
  {Volkov}}\ and\ \bibinfo {author} {\bibfnamefont {V.}~\bibnamefont
  {Akulov}},\ }\href {\doibase http://dx.doi.org/10.1016/0370-2693(73)90490-5}
  {\bibfield  {journal} {\bibinfo  {journal} {Phys. Lett. B}\ }\textbf
  {\bibinfo {volume} {46}},\ \bibinfo {pages} {109 } (\bibinfo {year}
  {1973})}\BibitemShut {NoStop}%
\bibitem [{\citenamefont {Fendley}\ \emph {et~al.}(2003)\citenamefont
  {Fendley}, \citenamefont {Schoutens},\ and\ \citenamefont
  {de~Boer}}]{PRL_Fendley_2003}%
  \BibitemOpen
  \bibfield  {author} {\bibinfo {author} {\bibfnamefont {P.}~\bibnamefont
  {Fendley}}, \bibinfo {author} {\bibfnamefont {K.}~\bibnamefont {Schoutens}},
  \ and\ \bibinfo {author} {\bibfnamefont {J.}~\bibnamefont {de~Boer}},\ }\href
  {\doibase 10.1103/PhysRevLett.90.120402} {\bibfield  {journal} {\bibinfo
  {journal} {Phys. Rev. Lett.}\ }\textbf {\bibinfo {volume} {90}},\ \bibinfo
  {pages} {120402} (\bibinfo {year} {2003})}\BibitemShut {NoStop}%
\bibitem [{\citenamefont {Fendley}\ and\ \citenamefont
  {Schoutens}(2005)}]{PRL_Fendley_2005}%
  \BibitemOpen
  \bibfield  {author} {\bibinfo {author} {\bibfnamefont {P.}~\bibnamefont
  {Fendley}}\ and\ \bibinfo {author} {\bibfnamefont {K.}~\bibnamefont
  {Schoutens}},\ }\href {\doibase 10.1103/PhysRevLett.95.046403} {\bibfield
  {journal} {\bibinfo  {journal} {Phys. Rev. Lett.}\ }\textbf {\bibinfo
  {volume} {95}},\ \bibinfo {pages} {046403} (\bibinfo {year}
  {2005})}\BibitemShut {NoStop}%
\bibitem [{\citenamefont {Yu}\ and\ \citenamefont {Yang}(2008)}]{PRL_Yu}%
  \BibitemOpen
  \bibfield  {author} {\bibinfo {author} {\bibfnamefont {Y.}~\bibnamefont
  {Yu}}\ and\ \bibinfo {author} {\bibfnamefont {K.}~\bibnamefont {Yang}},\
  }\href {\doibase 10.1103/PhysRevLett.100.090404} {\bibfield  {journal}
  {\bibinfo  {journal} {Phys. Rev. Lett.}\ }\textbf {\bibinfo {volume} {100}},\
  \bibinfo {pages} {090404} (\bibinfo {year} {2008})}\BibitemShut {NoStop}%
\bibitem [{\citenamefont {Huijse}\ \emph {et~al.}(2008)\citenamefont {Huijse},
  \citenamefont {Halverson}, \citenamefont {Fendley},\ and\ \citenamefont
  {Schoutens}}]{PRL_Huijse}%
  \BibitemOpen
  \bibfield  {author} {\bibinfo {author} {\bibfnamefont {L.}~\bibnamefont
  {Huijse}}, \bibinfo {author} {\bibfnamefont {J.}~\bibnamefont {Halverson}},
  \bibinfo {author} {\bibfnamefont {P.}~\bibnamefont {Fendley}}, \ and\
  \bibinfo {author} {\bibfnamefont {K.}~\bibnamefont {Schoutens}},\ }\href
  {\doibase 10.1103/PhysRevLett.101.146406} {\bibfield  {journal} {\bibinfo
  {journal} {Phys. Rev. Lett.}\ }\textbf {\bibinfo {volume} {101}},\ \bibinfo
  {pages} {146406} (\bibinfo {year} {2008})}\BibitemShut {NoStop}%
\bibitem [{\citenamefont {Huijse}\ and\ \citenamefont
  {Schoutens}(2008)}]{EPJB_huijse}%
  \BibitemOpen
  \bibfield  {author} {\bibinfo {author} {\bibfnamefont {L.}~\bibnamefont
  {Huijse}}\ and\ \bibinfo {author} {\bibfnamefont {K.}~\bibnamefont
  {Schoutens}},\ }\href@noop {} {\bibfield  {journal} {\bibinfo  {journal}
  {Eur. Phys. J. B}\ }\textbf {\bibinfo {volume} {64}},\ \bibinfo {pages} {543}
  (\bibinfo {year} {2008})}\BibitemShut {NoStop}%
\bibitem [{\citenamefont {Yu}\ and\ \citenamefont {Yang}(2010)}]{PRL_Yu_2010}%
  \BibitemOpen
  \bibfield  {author} {\bibinfo {author} {\bibfnamefont {Y.}~\bibnamefont
  {Yu}}\ and\ \bibinfo {author} {\bibfnamefont {K.}~\bibnamefont {Yang}},\
  }\href@noop {} {\bibfield  {journal} {\bibinfo  {journal} {Phys. Rev. Lett.}\
  }\textbf {\bibinfo {volume} {105}},\ \bibinfo {pages} {150605} (\bibinfo
  {year} {2010})}\BibitemShut {NoStop}%
\bibitem [{\citenamefont {Huijse}\ \emph {et~al.}(2012)\citenamefont {Huijse},
  \citenamefont {Mehta}, \citenamefont {Moran}, \citenamefont {Schoutens},\
  and\ \citenamefont {Vala}}]{NJP_Huijse}%
  \BibitemOpen
  \bibfield  {author} {\bibinfo {author} {\bibfnamefont {L.}~\bibnamefont
  {Huijse}}, \bibinfo {author} {\bibfnamefont {D.}~\bibnamefont {Mehta}},
  \bibinfo {author} {\bibfnamefont {N.}~\bibnamefont {Moran}}, \bibinfo
  {author} {\bibfnamefont {K.}~\bibnamefont {Schoutens}}, \ and\ \bibinfo
  {author} {\bibfnamefont {J.}~\bibnamefont {Vala}},\ }\href
  {http://stacks.iop.org/1367-2630/14/i=7/a=073002} {\bibfield  {journal}
  {\bibinfo  {journal} {New J. Phys.}\ }\textbf {\bibinfo {volume} {14}},\
  \bibinfo {pages} {073002} (\bibinfo {year} {2012})}\BibitemShut {NoStop}%
\bibitem [{\citenamefont {Lee}(2007)}]{PRB_Lee_2007}%
  \BibitemOpen
  \bibfield  {author} {\bibinfo {author} {\bibfnamefont {S.-S.}\ \bibnamefont
  {Lee}},\ }\href@noop {} {\bibfield  {journal} {\bibinfo  {journal} {Phys.
  Rev. B}\ }\textbf {\bibinfo {volume} {76}},\ \bibinfo {pages} {075103}
  (\bibinfo {year} {2007})}\BibitemShut {NoStop}%
\bibitem [{\citenamefont {Grover}\ \emph {et~al.}(2014)\citenamefont {Grover},
  \citenamefont {Sheng},\ and\ \citenamefont {Vishwanath}}]{Science_Grover}%
  \BibitemOpen
  \bibfield  {author} {\bibinfo {author} {\bibfnamefont {T.}~\bibnamefont
  {Grover}}, \bibinfo {author} {\bibfnamefont {D.~N.}\ \bibnamefont {Sheng}}, \
  and\ \bibinfo {author} {\bibfnamefont {A.}~\bibnamefont {Vishwanath}},\
  }\href {\doibase 10.1126/science.1248253} {\bibfield  {journal} {\bibinfo
  {journal} {Science}\ }\textbf {\bibinfo {volume} {344}},\ \bibinfo {pages}
  {280} (\bibinfo {year} {2014})}\BibitemShut {NoStop}%
\bibitem [{\citenamefont {Jian}\ \emph {et~al.}(2015)\citenamefont {Jian},
  \citenamefont {Jiang},\ and\ \citenamefont {Yao}}]{PRL_Jian}%
  \BibitemOpen
  \bibfield  {author} {\bibinfo {author} {\bibfnamefont {S.-K.}\ \bibnamefont
  {Jian}}, \bibinfo {author} {\bibfnamefont {Y.-F.}\ \bibnamefont {Jiang}}, \
  and\ \bibinfo {author} {\bibfnamefont {H.}~\bibnamefont {Yao}},\ }\href
  {\doibase 10.1103/PhysRevLett.114.237001} {\bibfield  {journal} {\bibinfo
  {journal} {Phys. Rev. Lett.}\ }\textbf {\bibinfo {volume} {114}},\ \bibinfo
  {pages} {237001} (\bibinfo {year} {2015})}\BibitemShut {NoStop}%
\bibitem [{\citenamefont {Rahmani}\ \emph {et~al.}(2015)\citenamefont
  {Rahmani}, \citenamefont {Zhu}, \citenamefont {Franz},\ and\ \citenamefont
  {Affleck}}]{PRL_Rahmani}%
  \BibitemOpen
  \bibfield  {author} {\bibinfo {author} {\bibfnamefont {A.}~\bibnamefont
  {Rahmani}}, \bibinfo {author} {\bibfnamefont {X.}~\bibnamefont {Zhu}},
  \bibinfo {author} {\bibfnamefont {M.}~\bibnamefont {Franz}}, \ and\ \bibinfo
  {author} {\bibfnamefont {I.}~\bibnamefont {Affleck}},\ }\href {\doibase
  10.1103/PhysRevLett.115.166401} {\bibfield  {journal} {\bibinfo  {journal}
  {Phys. Rev. Lett.}\ }\textbf {\bibinfo {volume} {115}},\ \bibinfo {pages}
  {166401} (\bibinfo {year} {2015})}\BibitemShut {NoStop}%
\bibitem [{\citenamefont {Li}\ \emph {et~al.}(2016)\citenamefont {Li},
  \citenamefont {Jiang},\ and\ \citenamefont {Yao}}]{arXiv_Li}%
  \BibitemOpen
  \bibfield  {author} {\bibinfo {author} {\bibfnamefont {Z.-X.}\ \bibnamefont
  {Li}}, \bibinfo {author} {\bibfnamefont {Y.-F.}\ \bibnamefont {Jiang}}, \
  and\ \bibinfo {author} {\bibfnamefont {H.}~\bibnamefont {Yao}},\ }\href@noop
  {} {\enquote {\bibinfo {title} {Edge quantum criticality and emergent
  supersymmetry in topological phases},}\ } (\bibinfo {year} {2016}),\ \Eprint
  {http://arxiv.org/abs/arXiv:1610.04616} {arXiv:1610.04616} \BibitemShut
  {NoStop}%
\bibitem [{\citenamefont {Jian}\ \emph {et~al.}(2016)\citenamefont {Jian},
  \citenamefont {Lin}, \citenamefont {Maciejko},\ and\ \citenamefont
  {Yao}}]{arXiv_Jian}%
  \BibitemOpen
  \bibfield  {author} {\bibinfo {author} {\bibfnamefont {S.-K.}\ \bibnamefont
  {Jian}}, \bibinfo {author} {\bibfnamefont {C.-H.}\ \bibnamefont {Lin}},
  \bibinfo {author} {\bibfnamefont {J.}~\bibnamefont {Maciejko}}, \ and\
  \bibinfo {author} {\bibfnamefont {H.}~\bibnamefont {Yao}},\ }\href@noop {}
  {\enquote {\bibinfo {title} {Emergence of supersymmetric quantum
  electrodynamics on the surface of a correlated topological insulator},}\ }
  (\bibinfo {year} {2016}),\ \Eprint {http://arxiv.org/abs/arXiv:1609.02146}
  {arXiv:1609.02146} \BibitemShut {NoStop}%
\bibitem [{\citenamefont {Nambu}\ and\ \citenamefont
  {Jona-Lasinio}(1961)}]{PR_Nambu}%
  \BibitemOpen
  \bibfield  {author} {\bibinfo {author} {\bibfnamefont {Y.}~\bibnamefont
  {Nambu}}\ and\ \bibinfo {author} {\bibfnamefont {G.}~\bibnamefont
  {Jona-Lasinio}},\ }\href {\doibase 10.1103/PhysRev.122.345} {\bibfield
  {journal} {\bibinfo  {journal} {Phys. Rev.}\ }\textbf {\bibinfo {volume}
  {122}},\ \bibinfo {pages} {345} (\bibinfo {year} {1961})}\BibitemShut
  {NoStop}%
\bibitem [{\citenamefont {Goldstone}(1961)}]{NC_Goldstone}%
  \BibitemOpen
  \bibfield  {author} {\bibinfo {author} {\bibfnamefont {J.}~\bibnamefont
  {Goldstone}},\ }\href {\doibase 10.1007/BF02812722} {\bibfield  {journal}
  {\bibinfo  {journal} {Nuovo Cimento}\ }\textbf {\bibinfo {volume} {19}},\
  \bibinfo {pages} {154} (\bibinfo {year} {1961})}\BibitemShut {NoStop}%
\bibitem [{\citenamefont {Goldstone}\ \emph {et~al.}(1962)\citenamefont
  {Goldstone}, \citenamefont {Salam},\ and\ \citenamefont
  {Weinberg}}]{PR_Goldstone}%
  \BibitemOpen
  \bibfield  {author} {\bibinfo {author} {\bibfnamefont {J.}~\bibnamefont
  {Goldstone}}, \bibinfo {author} {\bibfnamefont {A.}~\bibnamefont {Salam}}, \
  and\ \bibinfo {author} {\bibfnamefont {S.}~\bibnamefont {Weinberg}},\ }\href
  {\doibase 10.1103/PhysRev.127.965} {\bibfield  {journal} {\bibinfo  {journal}
  {Phys. Rev.}\ }\textbf {\bibinfo {volume} {127}},\ \bibinfo {pages} {965}
  (\bibinfo {year} {1962})}\BibitemShut {NoStop}%
\bibitem [{\citenamefont {Watanabe}\ and\ \citenamefont
  {Murayama}(2012)}]{PRL_Watanabe}%
  \BibitemOpen
  \bibfield  {author} {\bibinfo {author} {\bibfnamefont {H.}~\bibnamefont
  {Watanabe}}\ and\ \bibinfo {author} {\bibfnamefont {H.}~\bibnamefont
  {Murayama}},\ }\href {\doibase 10.1103/PhysRevLett.108.251602} {\bibfield
  {journal} {\bibinfo  {journal} {Phys. Rev. Lett.}\ }\textbf {\bibinfo
  {volume} {108}},\ \bibinfo {pages} {251602} (\bibinfo {year}
  {2012})}\BibitemShut {NoStop}%
\bibitem [{\citenamefont {Hidaka}(2013)}]{PRL_Hidaka}%
  \BibitemOpen
  \bibfield  {author} {\bibinfo {author} {\bibfnamefont {Y.}~\bibnamefont
  {Hidaka}},\ }\href {\doibase 10.1103/PhysRevLett.110.091601} {\bibfield
  {journal} {\bibinfo  {journal} {Phys. Rev. Lett.}\ }\textbf {\bibinfo
  {volume} {110}},\ \bibinfo {pages} {091601} (\bibinfo {year}
  {2013})}\BibitemShut {NoStop}%
\bibitem [{\citenamefont {Sannomiya}\ \emph {et~al.}(2016)\citenamefont
  {Sannomiya}, \citenamefont {Katsura},\ and\ \citenamefont
  {Nakayama}}]{U1Nicolai}%
  \BibitemOpen
  \bibfield  {author} {\bibinfo {author} {\bibfnamefont {N.}~\bibnamefont
  {Sannomiya}}, \bibinfo {author} {\bibfnamefont {H.}~\bibnamefont {Katsura}},
  \ and\ \bibinfo {author} {\bibfnamefont {Y.}~\bibnamefont {Nakayama}},\
  }\href {\doibase 10.1103/PhysRevD.94.045014} {\bibfield  {journal} {\bibinfo
  {journal} {Phys. Rev. D}\ }\textbf {\bibinfo {volume} {94}},\ \bibinfo
  {pages} {045014} (\bibinfo {year} {2016})}\BibitemShut {NoStop}%
\bibitem [{\citenamefont {Nicolai}(1976)}]{JPA_Nicolai_76}%
  \BibitemOpen
  \bibfield  {author} {\bibinfo {author} {\bibfnamefont {H.}~\bibnamefont
  {Nicolai}},\ }\href {http://stacks.iop.org/0305-4470/9/i=9/a=010} {\bibfield
  {journal} {\bibinfo  {journal} {J. Phys. A}\ }\textbf {\bibinfo {volume}
  {9}},\ \bibinfo {pages} {1497} (\bibinfo {year} {1976})}\BibitemShut
  {NoStop}%
\bibitem [{\citenamefont {Nicolai}(1977)}]{JPA_Nicolai_77}%
  \BibitemOpen
  \bibfield  {author} {\bibinfo {author} {\bibfnamefont {H.}~\bibnamefont
  {Nicolai}},\ }\href {http://stacks.iop.org/0305-4470/10/i=12/a=022}
  {\bibfield  {journal} {\bibinfo  {journal} {J. Phys. A}\ }\textbf {\bibinfo
  {volume} {10}},\ \bibinfo {pages} {2143} (\bibinfo {year}
  {1977})}\BibitemShut {NoStop}%
\bibitem [{\citenamefont {Heikkil{\"a}}\ and\ \citenamefont
  {Volovik}(2010)}]{JETP_Volovik}%
  \BibitemOpen
  \bibfield  {author} {\bibinfo {author} {\bibfnamefont {T.~T.}\ \bibnamefont
  {Heikkil{\"a}}}\ and\ \bibinfo {author} {\bibfnamefont {G.~E.}\ \bibnamefont
  {Volovik}},\ }\href {\doibase 10.1134/S0021364010220091} {\bibfield
  {journal} {\bibinfo  {journal} {JETP Letters}\ }\textbf {\bibinfo {volume}
  {92}},\ \bibinfo {pages} {681} (\bibinfo {year} {2010})}\BibitemShut
  {NoStop}%
\bibitem [{\citenamefont {Guinea}\ \emph {et~al.}(2006)\citenamefont {Guinea},
  \citenamefont {Castro~Neto},\ and\ \citenamefont {Peres}}]{PRB_Guinea}%
  \BibitemOpen
  \bibfield  {author} {\bibinfo {author} {\bibfnamefont {F.}~\bibnamefont
  {Guinea}}, \bibinfo {author} {\bibfnamefont {A.~H.}\ \bibnamefont
  {Castro~Neto}}, \ and\ \bibinfo {author} {\bibfnamefont {N.~M.~R.}\
  \bibnamefont {Peres}},\ }\href {\doibase 10.1103/PhysRevB.73.245426}
  {\bibfield  {journal} {\bibinfo  {journal} {Phys. Rev. B}\ }\textbf {\bibinfo
  {volume} {73}},\ \bibinfo {pages} {245426} (\bibinfo {year}
  {2006})}\BibitemShut {NoStop}%
\bibitem [{\citenamefont {Liu}\ and\ \citenamefont
  {Zunger}(2016)}]{Zunger_arxiv}%
  \BibitemOpen
  \bibfield  {author} {\bibinfo {author} {\bibfnamefont {Q.}~\bibnamefont
  {Liu}}\ and\ \bibinfo {author} {\bibfnamefont {A.}~\bibnamefont {Zunger}},\
  }\href@noop {} {\enquote {\bibinfo {title} {Cubic dirac fermion in
  quasi-one-dimensional transition-metal mono-chalcogenide},}\ } (\bibinfo
  {year} {2016}),\ \Eprint {http://arxiv.org/abs/arXiv:1611.04147}
  {arXiv:1611.04147} \BibitemShut {NoStop}%
\bibitem [{\citenamefont {Biswas}\ \emph {et~al.}(2011)\citenamefont {Biswas},
  \citenamefont {Fu}, \citenamefont {Laumann},\ and\ \citenamefont
  {Sachdev}}]{PRB_Biswas}%
  \BibitemOpen
  \bibfield  {author} {\bibinfo {author} {\bibfnamefont {R.~R.}\ \bibnamefont
  {Biswas}}, \bibinfo {author} {\bibfnamefont {L.}~\bibnamefont {Fu}}, \bibinfo
  {author} {\bibfnamefont {C.~R.}\ \bibnamefont {Laumann}}, \ and\ \bibinfo
  {author} {\bibfnamefont {S.}~\bibnamefont {Sachdev}},\ }\href {\doibase
  10.1103/PhysRevB.83.245131} {\bibfield  {journal} {\bibinfo  {journal} {Phys.
  Rev. B}\ }\textbf {\bibinfo {volume} {83}},\ \bibinfo {pages} {245131}
  (\bibinfo {year} {2011})}\BibitemShut {NoStop}%
\bibitem [{\citenamefont {Wang}(1992)}]{PRB_Wang}%
  \BibitemOpen
  \bibfield  {author} {\bibinfo {author} {\bibfnamefont {Y.~R.}\ \bibnamefont
  {Wang}},\ }\href {\doibase 10.1103/PhysRevB.45.12604} {\bibfield  {journal}
  {\bibinfo  {journal} {Phys. Rev. B}\ }\textbf {\bibinfo {volume} {45}},\
  \bibinfo {pages} {12604} (\bibinfo {year} {1992})}\BibitemShut {NoStop}%
\bibitem [{\citenamefont {Gross}\ and\ \citenamefont
  {Rosenhaus}(2016)}]{Gross_arXiv}%
  \BibitemOpen
  \bibfield  {author} {\bibinfo {author} {\bibfnamefont {D.~J.}\ \bibnamefont
  {Gross}}\ and\ \bibinfo {author} {\bibfnamefont {V.}~\bibnamefont
  {Rosenhaus}},\ }\href@noop {} {\enquote {\bibinfo {title} {A generalization
  of {Sachdev-Ye-Kitaev}},}\ } (\bibinfo {year} {2016}),\ \Eprint
  {http://arxiv.org/abs/arXiv:1610.01569} {arXiv:1610.01569} \BibitemShut
  {NoStop}%
\bibitem [{\citenamefont {Fu}\ \emph {et~al.}(2017)\citenamefont {Fu},
  \citenamefont {Gaiotto}, \citenamefont {Maldacena},\ and\ \citenamefont
  {Sachdev}}]{Maldacena_arXiv}%
  \BibitemOpen
  \bibfield  {author} {\bibinfo {author} {\bibfnamefont {W.}~\bibnamefont
  {Fu}}, \bibinfo {author} {\bibfnamefont {D.}~\bibnamefont {Gaiotto}},
  \bibinfo {author} {\bibfnamefont {J.}~\bibnamefont {Maldacena}}, \ and\
  \bibinfo {author} {\bibfnamefont {S.}~\bibnamefont {Sachdev}},\ }\href
  {\doibase 10.1103/PhysRevD.95.026009} {\bibfield  {journal} {\bibinfo
  {journal} {Phys. Rev. D}\ }\textbf {\bibinfo {volume} {95}},\ \bibinfo
  {pages} {026009} (\bibinfo {year} {2017})}\BibitemShut {NoStop}%
\bibitem [{\citenamefont {de~Gier}\ \emph {et~al.}(2016)\citenamefont
  {de~Gier}, \citenamefont {Feher}, \citenamefont {Nienhuis},\ and\
  \citenamefont {Rusaczonek}}]{JSM_Gier}%
  \BibitemOpen
  \bibfield  {author} {\bibinfo {author} {\bibfnamefont {J.}~\bibnamefont
  {de~Gier}}, \bibinfo {author} {\bibfnamefont {G.~Z.}\ \bibnamefont {Feher}},
  \bibinfo {author} {\bibfnamefont {B.}~\bibnamefont {Nienhuis}}, \ and\
  \bibinfo {author} {\bibfnamefont {M.}~\bibnamefont {Rusaczonek}},\ }\href
  {http://stacks.iop.org/1742-5468/2016/i=2/a=023104} {\bibfield  {journal}
  {\bibinfo  {journal} {J. Stat. Mech.}\ ,\ \bibinfo {pages} {023104}}
  (\bibinfo {year} {2016})}\BibitemShut {NoStop}%
\bibitem [{\citenamefont {Essler}\ \emph {et~al.}(2005)\citenamefont {Essler},
  \citenamefont {Frahm}, \citenamefont {G{\" o}hmann}, \citenamefont {Kl{\"
  u}mper},\ and\ \citenamefont {Korepin}}]{Essler_Hubbard}%
  \BibitemOpen
  \bibfield  {author} {\bibinfo {author} {\bibfnamefont {F.~H.~L.}\
  \bibnamefont {Essler}}, \bibinfo {author} {\bibfnamefont {H.}~\bibnamefont
  {Frahm}}, \bibinfo {author} {\bibfnamefont {F.}~\bibnamefont {G{\" o}hmann}},
  \bibinfo {author} {\bibfnamefont {A.}~\bibnamefont {Kl{\" u}mper}}, \ and\
  \bibinfo {author} {\bibfnamefont {V.~E.}\ \bibnamefont {Korepin}},\ }\href
  {\doibase 10.1017/CBO9780511534843} {\emph {\bibinfo {title} {The
  One-Dimensional Hubbard Model:}}}\ (\bibinfo  {publisher} {Cambridge
  University Press},\ \bibinfo {address} {Cambridge},\ \bibinfo {year}
  {2005})\BibitemShut {NoStop}%
\bibitem [{\citenamefont {Moriya}(2016)}]{Moriya_arXiv}%
  \BibitemOpen
  \bibfield  {author} {\bibinfo {author} {\bibfnamefont {H.}~\bibnamefont
  {Moriya}},\ }\href@noop {} {\enquote {\bibinfo {title} {Breakdown of
  ergodicity induced by infinitely many local kinematical supercharges for the
  nicolai supersymmetric fermion lattice model},}\ } (\bibinfo {year} {2016}),\
  \Eprint {http://arxiv.org/abs/arXiv:1610.09142} {arXiv:1610.09142}
  \BibitemShut {NoStop}%
\bibitem [{\citenamefont {Anderson}(1951)}]{PR_Anderson}%
  \BibitemOpen
  \bibfield  {author} {\bibinfo {author} {\bibfnamefont {P.~W.}\ \bibnamefont
  {Anderson}},\ }\href {\doibase 10.1103/PhysRev.83.1260} {\bibfield  {journal}
  {\bibinfo  {journal} {Phys. Rev.}\ }\textbf {\bibinfo {volume} {83}},\
  \bibinfo {pages} {1260} (\bibinfo {year} {1951})}\BibitemShut {NoStop}%
\bibitem [{\citenamefont {Valent\'{\i}}\ \emph {et~al.}(1991)\citenamefont
  {Valent\'{\i}}, \citenamefont {Hirschfeld},\ and\ \citenamefont
  {Angl\'es~d'Auriac}}]{PRB_Valent}%
  \BibitemOpen
  \bibfield  {author} {\bibinfo {author} {\bibfnamefont {R.}~\bibnamefont
  {Valent\'{\i}}}, \bibinfo {author} {\bibfnamefont {P.~J.}\ \bibnamefont
  {Hirschfeld}}, \ and\ \bibinfo {author} {\bibfnamefont {J.~C.}\ \bibnamefont
  {Angl\'es~d'Auriac}},\ }\href {\doibase 10.1103/PhysRevB.44.3995} {\bibfield
  {journal} {\bibinfo  {journal} {Phys. Rev. B}\ }\textbf {\bibinfo {volume}
  {44}},\ \bibinfo {pages} {3995} (\bibinfo {year} {1991})}\BibitemShut
  {NoStop}%
\bibitem [{\citenamefont {Nie}\ \emph {et~al.}(2013)\citenamefont {Nie},
  \citenamefont {Katsura},\ and\ \citenamefont {Oshikawa}}]{PRL_Nie}%
  \BibitemOpen
  \bibfield  {author} {\bibinfo {author} {\bibfnamefont {W.}~\bibnamefont
  {Nie}}, \bibinfo {author} {\bibfnamefont {H.}~\bibnamefont {Katsura}}, \ and\
  \bibinfo {author} {\bibfnamefont {M.}~\bibnamefont {Oshikawa}},\ }\href
  {\doibase 10.1103/PhysRevLett.111.100402} {\bibfield  {journal} {\bibinfo
  {journal} {Phys. Rev. Lett.}\ }\textbf {\bibinfo {volume} {111}},\ \bibinfo
  {pages} {100402} (\bibinfo {year} {2013})}\BibitemShut {NoStop}%
\bibitem [{\citenamefont {Nie}\ \emph {et~al.}(2014)\citenamefont {Nie},
  \citenamefont {Katsura},\ and\ \citenamefont {Oshikawa}}]{Nie_arXiv}%
  \BibitemOpen
  \bibfield  {author} {\bibinfo {author} {\bibfnamefont {W.}~\bibnamefont
  {Nie}}, \bibinfo {author} {\bibfnamefont {H.}~\bibnamefont {Katsura}}, \ and\
  \bibinfo {author} {\bibfnamefont {M.}~\bibnamefont {Oshikawa}},\ }\href@noop
  {} {\enquote {\bibinfo {title} {Particle statistics, frustration, and
  ground-state energy},}\ } (\bibinfo {year} {2014}),\ \Eprint
  {http://arxiv.org/abs/arXiv:1401.2090} {arXiv:1401.2090} \BibitemShut
  {NoStop}%
\bibitem [{\citenamefont {Beccaria}\ \emph {et~al.}(2004)\citenamefont
  {Beccaria}, \citenamefont {De~Angelis}, \citenamefont {Campostrini},\ and\
  \citenamefont {Feo}}]{PRD_Beccaria}%
  \BibitemOpen
  \bibfield  {author} {\bibinfo {author} {\bibfnamefont {M.}~\bibnamefont
  {Beccaria}}, \bibinfo {author} {\bibfnamefont {G.~F.}\ \bibnamefont
  {De~Angelis}}, \bibinfo {author} {\bibfnamefont {M.}~\bibnamefont
  {Campostrini}}, \ and\ \bibinfo {author} {\bibfnamefont {A.}~\bibnamefont
  {Feo}},\ }\href {\doibase 10.1103/PhysRevD.70.035011} {\bibfield  {journal}
  {\bibinfo  {journal} {Phys. Rev. D}\ }\textbf {\bibinfo {volume} {70}},\
  \bibinfo {pages} {035011} (\bibinfo {year} {2004})}\BibitemShut {NoStop}%
\bibitem [{\citenamefont {Feynman}(1954)}]{PR_Feynman}%
  \BibitemOpen
  \bibfield  {author} {\bibinfo {author} {\bibfnamefont {R.~P.}\ \bibnamefont
  {Feynman}},\ }\href {\doibase 10.1103/PhysRev.94.262} {\bibfield  {journal}
  {\bibinfo  {journal} {Phys. Rev.}\ }\textbf {\bibinfo {volume} {94}},\
  \bibinfo {pages} {262} (\bibinfo {year} {1954})}\BibitemShut {NoStop}%
\bibitem [{\citenamefont {Horsch}\ and\ \citenamefont {von~der
  Linden}(1988)}]{ZPB_Horsch}%
  \BibitemOpen
  \bibfield  {author} {\bibinfo {author} {\bibfnamefont {P.}~\bibnamefont
  {Horsch}}\ and\ \bibinfo {author} {\bibfnamefont {W.}~\bibnamefont {von~der
  Linden}},\ }\href {\doibase 10.1007/BF01312134} {\bibfield  {journal}
  {\bibinfo  {journal} {Z. Phys. B}\ }\textbf {\bibinfo {volume} {72}},\
  \bibinfo {pages} {181} (\bibinfo {year} {1988})}\BibitemShut {NoStop}%
\bibitem [{\citenamefont {Stringari}(1994)}]{PRB_Stringari}%
  \BibitemOpen
  \bibfield  {author} {\bibinfo {author} {\bibfnamefont {S.}~\bibnamefont
  {Stringari}},\ }\href {\doibase 10.1103/PhysRevB.49.6710} {\bibfield
  {journal} {\bibinfo  {journal} {Phys. Rev. B}\ }\textbf {\bibinfo {volume}
  {49}},\ \bibinfo {pages} {6710} (\bibinfo {year} {1994})}\BibitemShut
  {NoStop}%
\bibitem [{\citenamefont {Momoi}(1994)}]{JPSJ_Momoi}%
  \BibitemOpen
  \bibfield  {author} {\bibinfo {author} {\bibfnamefont {T.}~\bibnamefont
  {Momoi}},\ }\href {\doibase 10.1143/JPSJ.63.2507} {\bibfield  {journal}
  {\bibinfo  {journal} {J. Phys. Soc. Jpn.}\ }\textbf {\bibinfo {volume}
  {63}},\ \bibinfo {pages} {2507} (\bibinfo {year} {1994})}\BibitemShut
  {NoStop}%
\bibitem [{Note1()}]{Note1}%
  \BibitemOpen
  \bibinfo {note} {From numerical results, the eigenvalues of the translation
  operator are $\pm 1$ in the ground states (see Figs. \ref {fig:dispO} and
  \ref {fig:dispE}), which means that trial states are orthogonal to all the
  ground states when $p$ is finite and small enough.}\BibitemShut {Stop}%
\bibitem [{\citenamefont {Pitaevskii}\ and\ \citenamefont
  {Stringari}(1991)}]{JLTP_Pitaevskii}%
  \BibitemOpen
  \bibfield  {author} {\bibinfo {author} {\bibfnamefont {L.}~\bibnamefont
  {Pitaevskii}}\ and\ \bibinfo {author} {\bibfnamefont {S.}~\bibnamefont
  {Stringari}},\ }\href {\doibase 10.1007/BF00682193} {\bibfield  {journal}
  {\bibinfo  {journal} {J. Low. Temp. Phys.}\ }\textbf {\bibinfo {volume}
  {85}},\ \bibinfo {pages} {377} (\bibinfo {year} {1991})}\BibitemShut
  {NoStop}%
\bibitem [{\citenamefont {Haldane}(1982)}]{Haldane_1982}%
  \BibitemOpen
  \bibfield  {author} {\bibinfo {author} {\bibfnamefont {F.~D.~M.}\
  \bibnamefont {Haldane}},\ }\href
  {http://stacks.iop.org/0022-3719/15/i=36/a=008} {\bibfield  {journal}
  {\bibinfo  {journal} {J. Phys. C}\ }\textbf {\bibinfo {volume} {15}},\
  \bibinfo {pages} {L1309} (\bibinfo {year} {1982})}\BibitemShut {NoStop}%
\bibitem [{Note2()}]{Note2}%
  \BibitemOpen
  \bibinfo {note} {The On-Line Encyclopedia of Integer Sequences,
  http://oeis.org}\BibitemShut {NoStop}%
\bibitem [{Note3()}]{Note3}%
  \BibitemOpen
  \bibinfo {note} {See also \cite {Gross_arXiv}. Our result was communicated to
  V.~Rosenhaus by one of the authors (Y.~N.) at Strings 2016.}\BibitemShut
  {Stop}%
\end{thebibliography}%

\end{document}